\documentclass[format=acmsmall,table,nonacm]{acmart}
\usepackage{blindtext, graphicx}
\usepackage{csquotes}
\usepackage{soul}
\usepackage{enumerate}
\usepackage{diagbox}
\usepackage{multirow}
\usepackage{listings}
\usepackage{longtable}
\usepackage{tcolorbox}
\usepackage{hhline}
\usepackage{tikz}
\newcommand*\circled[1]{\tikz[baseline=(char.base)]{
            \node[shape=circle,fill,inner sep=2pt] (char) {\textcolor{white}{#1}};}}

\newif\iffull
\fullfalse

\newcommand{\tabitem}{~~\llap{\textbullet}~~}

\usepackage{mathtools}
\usepackage{bbold}

\newcolumntype{P}[1]{>{\centering\arraybackslash}m{#1}}

\usepackage{makecell}

\acmJournal{CSUR}

\setcopyright{none}
\renewcommand\footnotetextcopyrightpermission[1]{}

\begin{document}

\title[A Survey on Data-driven Software Vulnerability Assessment and Prioritization]{A Survey on Data-driven Software Vulnerability Assessment and Prioritization}

\author{Triet H. M. Le}
\email{triet.h.le@adelaide.edu.au}
\author{Huaming Chen}
\email{huaming.chen@adelaide.edu.au}
\affiliation{
\institution{CREST - The Centre for Research on Engineering Software Technologies, The University of Adelaide}
\city{Adelaide}
\country{Australia}
}

\author{M. Ali Babar}
\email{ali.babar@adelaide.edu.au}
\affiliation{
\institution{CREST - The Centre for Research on Engineering Software Technologies, The University of Adelaide}
\city{Adelaide}
\country{Australia}
}
\affiliation{\institution{Cyber Security Cooperative Research Centre}
\city{}
\country{Australia}}

\begin{abstract}

Software Vulnerabilities (SVs) are increasing in complexity and scale, posing great security risks to many software systems. Given the limited resources in practice, SV assessment and prioritization help practitioners devise optimal SV mitigation plans based on various SV characteristics. The surges in SV data sources and data-driven techniques such as Machine Learning and Deep Learning have taken SV assessment and prioritization to the next level. Our survey provides a taxonomy of the past research efforts and highlights the best practices for data-driven SV assessment and prioritization. We also discuss the current limitations and propose potential solutions to address such issues.

\end{abstract}



\authorsaddresses{
Authors' addresses: Triet H. M. Le, triet.h.le@adelaide.edu.au; Huaming Chen, huaming.chen@adelaide.edu.au; M. Ali Babar, ali.babar@adelaide.edu.au,  The University of Adelaide, Adelaide, Australia.}

\maketitle

\lstset{language=Java,keywordstyle={\bfseries}}

\section{Introduction}\label{sec:introduction}
Software Vulnerabilities (SVs) can negatively affect the confidentiality, integrity and availability of software systems~\cite{ghaffarian2017software}.
The exploitation of these SVs such as the Heartbleed attack\footnote{\url{https://heartbleed.com}} can damage the operations and reputation of millions of systems and organizations globally, resulting in huge financial losses as well. Therefore, it is important to remediate critical SVs as promptly as possible.

Vulnerability assessment is required to prioritize the remediation of critical SVs among a large and increasing number of SVs each year~\cite{smyth2017software} (e.g., more than 20,000 SVs were reported on National Vulnerability Database (NVD)~\cite{nvd} in 2021). SV assessment includes tasks that determine various characteristics such as the types, exploitability, impact and severity levels of SVs~\cite{spanos2018multi,le2019automated,duan2021automated}. Such characteristics help understand and select high-priority SVs to resolve early given the limited effort and resources. For example, an identified cross-site scripting (XSS) or SQL injection vulnerability in a web application will likely require an urgent remediation plan. These two types of SVs are well-known and can be easily exploited by attackers to gain unauthorized access and compromise sensitive data/information. On the other hand, an SV that requires admin access or happens only in a local network will probably have a lower priority since only a few people can initiate an attack.

There has been an active research area to assess and prioritize SVs using increasingly large data from multiple sources.
Many studies in this area have proposed different Natural Language Processing (NLP), Machine Learning (ML) and Deep Learning (DL) techniques to leverage such data to automate various tasks such as predicting the Common Vulnerability Scoring System (CVSS)~\cite{cvss} metrics (e.g.,~\cite{le2019automated,spanos2018multi,han2017learning}) or public exploits (e.g.,~\cite{bozorgi2010beyond,sabottke2015vulnerability,bullough2017predicting}).
These prediction models can learn the patterns automatically from vast SV data, which would be otherwise impossible to do manually. Such patterns are utilized to speed up the assessment and prioritization processes of ever-increasing and more complex SVs, significantly reducing practitioners' effort.
Despite the rising research interest in data-driven SV assessment and prioritization, to the best of our knowledge, there has been no comprehensive survey on the state-of-the-art methods and existing challenges in this area.

\begin{table}[!t]
\fontsize{8}{9}\selectfont
\caption{Comparison of contributions between our survey and the existing related surveys/reviews.}

\label{tab:survey_comparison}
\centering
\begin{tabular}{|l|P{3cm}|P{3cm}|P{3cm}|}
\hline
\multicolumn{1}{|l|}{\diagbox[height=1.3cm, width=3.6cm]{\raisebox{2\height}{\enspace \textbf{Study}}}{\raisebox{-1\height}{\enspace \textbf{Contribution}}}} &
\multicolumn{1}{P{2.9cm}|}{\centering\textbf{Focus on SV assessment \& prioritization}} & \multicolumn{1}{P{2.9cm}|}{\centering\textbf{Analysis of SV\\data sources}} & \multicolumn{1}{P{2.9cm}|}{\centering\textbf{Analysis of data-\\driven approaches (NLP/ML/DL)}}
\\
\hline
\makecell[l]{Ghaffarian et al. 2017~\cite{ghaffarian2017software}} & \multicolumn{1}{c|}{--} & \multicolumn{1}{c|}{--} & \makecell[c]{\checkmark (Mostly ML)} \\
\hline
\makecell[l]{Lin et al. 2020~\cite{lin2020software}\\ Semasaba et al. 2020~\cite{semasaba2020literature}\\ Singh et al. 2020~\cite{singh2020applying}\\ Zeng et al. 2020~\cite{zeng2020software}} & \multicolumn{1}{c|}{--} & \multicolumn{1}{c|}{--} & \makecell[c]{\checkmark (Mostly DL)} \\
\hline
\multicolumn{1}{|p{2.8cm}|}{Pastor et al. 2020~\cite{pastor2020not}} & \multicolumn{1}{c|}{--} & \makecell[c]{\checkmark (OSINT)} & \multicolumn{1}{c|}{--} \\
\hline
\makecell[l]{Sun et al. 2018~\cite{sun2018data}\\ Evangelista et al. 2020~\cite{evangelista2020systematic}} & \multicolumn{1}{c|}{--} & \makecell[c]{\checkmark (OSINT)} & \multicolumn{1}{c|}{\checkmark} \\
\hline
\makecell[l]{Khan et al. 2018~\cite{khan2018review}} & \makecell[c]{\checkmark (Rule-based methods)} & \multicolumn{1}{c|}{--} & \multicolumn{1}{c|}{--} \\
\hline
\multicolumn{1}{|p{2.8cm}|}{Kritikos et al. 2019~\cite{kritikos2019survey}} & \makecell[c]{\checkmark (Static analysis)} & \multicolumn{1}{c|}{\checkmark} & \multicolumn{1}{c|}{--} \\
\hline
\multicolumn{1}{|p{3.2cm}|}{Dissanayake et al. 2020~\cite{dissanayake2020software}} & \multicolumn{1}{c|}{\checkmark (Socio-technical aspects)} & \multicolumn{1}{c|}{--} & \multicolumn{1}{c|}{--} \\
\hline\hline
\rowcolor{lightgray}
\multicolumn{1}{|l|}{\textbf{Our survey}} & \multicolumn{1}{c|}{\textbf{\checkmark}} & \multicolumn{1}{c|}{\textbf{\checkmark}} & \multicolumn{1}{c|}{\textbf{\checkmark}} \\
\hline
\end{tabular}
\end{table}

\noindent \textbf{Our Contributions}. \circled{1} We are the first to review in-depth the research studies that automate \textit{data-driven SV assessment and prioritization} tasks leveraging SV data and NLP/ML/DL techniques.
\circled{2} We categorize and describe the key tasks performed in relevant primary studies.
\circled{3} We synthesize and discuss the pros and cons of data, features, models, evaluation methods and metrics commonly used in the reviewed studies.
\circled{4} We highlight the challenges with the current practices and propose potential solutions moving forward.
Our findings can provide useful guidelines for researchers and practitioners to effectively utilize data to perform SV assessment and prioritization.
An online and up-to-date (by accepting external contributions) version of the survey can be found at \textcolor{blue}{\url{https://github.com/lhmtriet/awesome-vulnerability-assessment}}.

\noindent \textbf{Related Work}.
There have been several existing surveys/reviews on SV analysis and prediction, but they are fundamentally different from ours (see Table~\ref{tab:survey_comparison}).
Ghaffarian et al.~\cite{ghaffarian2017software} conducted a seminal survey on ML-based SV analysis and discovery.
Subsequently, several studies~\cite{zeng2020software,singh2020applying,semasaba2020literature,lin2020software} reviewed DL techniques for detecting vulnerable code.
However, these prior reviews did not describe how ML/DL techniques can be used to assess and prioritize the detected SVs.
There have been other relevant reviews on using Open Source Intelligence (OSINT) (e.g., phishing or malicious emails/URLs/IPs) to make informed security decisions~\cite{pastor2020not,evangelista2020systematic,sun2018data}. However, these OSINT reviews did not explicitly discuss the use of SV data and how such data can be leveraged to automate the assessment and prioritization processes. Moreover, most of the reviews on SV assessment and prioritization have focused on either static analysis tools~\cite{kritikos2019survey} or rule-based approaches (e.g., expert systems or ontologies)~\cite{khan2018review}.
These methods rely on pre-defined patterns and struggle to work with new types and different data sources of SVs compared to contemporary ML or DL approaches presented in our survey~\cite{han2011data,goodfellow2016deep}.
Recently, Dissanayake et al.~\cite{dissanayake2020software} reviewed the socio-technical challenges and solutions for security patch management that involves SV assessment and prioritization after SV patches are identified.
Unlike~\cite{dissanayake2020software}, we focus on the challenges, solutions and practices of automating various SV assessment and prioritization tasks with data-driven techniques.
We also consider all types of SV assessment/prioritization regardless of the patch availability.

\noindent \textbf{Paper Outline}. The rest of the paper is organized as follows. Section~\ref{sec:survey_overview} presents the scope, methodology and taxonomy covered in this survey. Sections~\ref{sec:exploit_prediction},~\ref{sec:impact_prediction},~\ref{sec:severity_prediction},~\ref{sec:type_prediction} and ~\ref{sec:miscellaneous} review the studies in each theme of the taxonomy. Section~\ref{sec:elements_analysis} identifies and discusses the common practices and respective implications for data-driven SV assessment and prioritization.
Section~\ref{sec:challenges} discusses the open challenges and proposes some future directions of this research area. Finally, section~\ref{sec:conclusions} concludes the survey.

\section{Survey Overview}
\label{sec:survey_overview}

\subsection{Background and Scope of the Survey}\label{subsec:scope}

Our survey's focus is on \textit{data-driven SV assessment and prioritization}. The \textit{assessment} and \textit{prioritization} phases are between the SV \textit{discovery}/\textit{detection} and SV \textit{remediation}/\textit{mitigation}/\textit{fixing}/\textit{patching} phases in an SV management lifecycle~\cite{foreman2019vulnerability}. The \textit{assessment} phase unveils the characteristics of the SVs found in the \textit{discovery} phase to locate ``hot spots'' that contain many highly critical/severe SVs and require higher attention in a system. In the \textit{prioritization} phase, practitioners use the assessment outputs to devise an optimal remediation plan, i.e., the order/priority of fixing each SV, based on available human and technological resources. SVs would then be mitigated/fixed accordingly to the prioritized plan in the \textit{remediation} phase.
Unlike the existing surveys on rule-based or experience-based SV assessment and prioritization~\cite{kritikos2019survey,khan2018review,dissanayake2020software} that hardly utilize the potential of SV data in the wild, this survey aims to review research papers that have leveraged such data to automate tasks in this area using data-driven models.
To keep our focus, we do not consider papers that only perform manual analyses or descriptive statistics (e.g., taking mean/median/variation of data) without using any data-driven models as these techniques cannot automatically assess or prioritize new SVs.
We also do not directly compare the absolute performance of all the related studies as they did not use exactly the same experimental setup (e.g., data sources and model configurations). While it is theoretically possible to perform a comparative evaluation of the identified techniques by establishing and using a common setup, this type of evaluation is out of the scope of this survey.
However, we still cover the key directions/techniques of the studies in sections~\mbox{\ref{sec:exploit_prediction}},~\mbox{\ref{sec:impact_prediction}},~\mbox{\ref{sec:severity_prediction}},~\mbox{\ref{sec:type_prediction}} and~\mbox{\ref{sec:miscellaneous}}.
We also provide in-depth discussion on the common practices and challenges of these studies and suggest some potential directions to advance the field in sections~\mbox{\ref{sec:elements_analysis}} and \mbox{\ref{sec:challenges}}.

\begin{figure*}[t]
  \centering
  \includegraphics[width=0.69\linewidth,keepaspectratio]{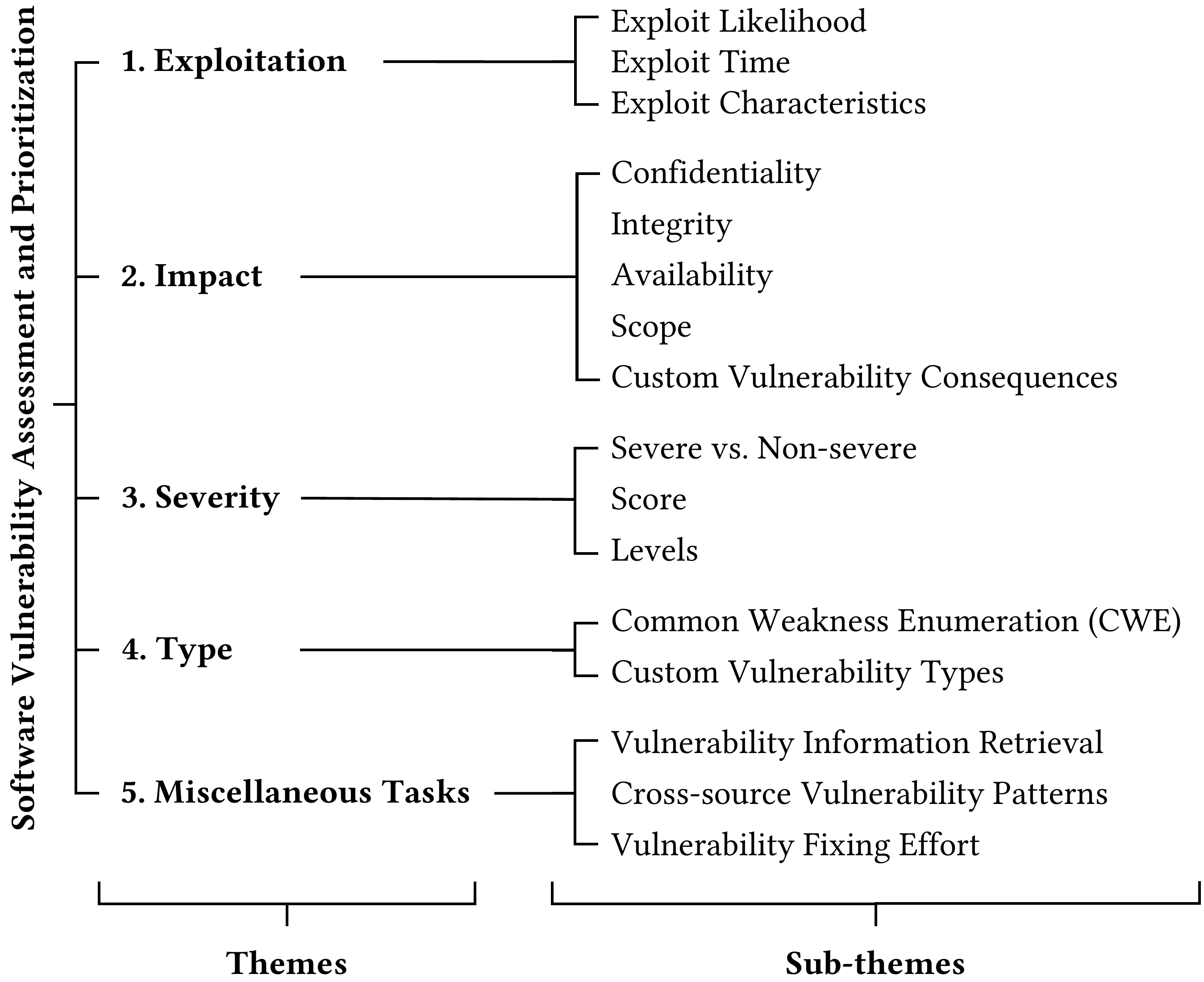}

  \caption{Taxonomy of studies on data-driven SV assessment and prioritization.}
  \label{fig:taxonomy}
\end{figure*}

\subsection{Methodology}
\label{subsec:methodology}
\noindent \textbf{Study selection}. Our study selection was inspired by the Systematic Literature Review guidelines~\mbox{\cite{keele2007guidelines}}. Due to the space limit, we only included the key steps of our selection here. The full details can be found at~\mbox{\cite{supp_materials}}. We first designed the search string: ``\textit{`software' AND vulner* AND (learn* OR data* OR predict*) AND (priority* OR assess* OR impact* OR exploit* OR severity*) AND NOT (fuzz* OR dynamic* OR intrusion OR adversari* OR malware* OR `vulnerability detection' OR `vulnerability discovery' OR `vulnerability identification' OR `vulnerability prediction')}''. This search string covered the key papers (i.e., with more than 50 citations) in the area and excluded many papers on general security and SV detection. We then adapted this string\footnote{This search string was customized for each database and the database-wise search strings can be found at~\mbox{\cite{supp_materials}}.} to retrieve an initial list of 1,765 papers up to April 2021 from various commonly used databases such as IEEE Xplore, ACM Digital Library, Scopus, SpringerLink and Wiley. We also defined the inclusion/exclusion criteria (see~\mbox{\cite{supp_materials}} for more details) to filter out irrelevant/low-quality studies with respect to our scope in section~\mbox{\ref{subsec:scope}}. Based on these criteria and the titles and abstracts and keywords of 1,765 initial papers, we removed 1,550 papers. After reading the full-text and applying the criteria on the remaining 215 papers, we obtained 70 papers directly related to data-driven SV assessment and prioritization.
To further increase the coverage of studies, we performed backward and forward snowballing on these 70 papers (using the above sources and Google Scholar) and identified 14 more papers that satisfied the inclusion/exclusion criteria.
In total, we included 84 studies for our survey. We do not claim that we have collected all the papers in this area, but we believe that our selection covered most of the key studies to unveil the practices of data-driven SV assessment and prioritization.

\noindent \textbf{Data extraction and synthesis of the selected studies}. We followed the steps of thematic analysis~\mbox{\cite{cruzes2011research}} to identify the taxonomy of data-driven SV assessment and prioritization tasks in sections~\mbox{\ref{sec:exploit_prediction}},~\mbox{\ref{sec:impact_prediction}},~\mbox{\ref{sec:severity_prediction}},~\mbox{\ref{sec:type_prediction}} and~\mbox{\ref{sec:miscellaneous}} as well as the key practices of data-driven model building for automating these tasks in section~\mbox{\ref{sec:elements_analysis}}. We first conducted a pilot study of 20 papers to familiarize ourselves with data to be extracted from the primary studies. After that, we generated initial codes and then merged them iteratively in several rounds to create themes. Two of the authors performed the analysis independently, in which each author analyzed half of the selected papers and then reviewed the analysis output of the other author. Any disagreements were resolved through discussions.

\subsection{Taxonomy of Data-driven Software Vulnerability Assessment and Prioritization}\label{subsec:taxonomy}

Based on the scope in section~\ref{subsec:scope} and the methodology in section~\ref{subsec:methodology}, we identified five main themes of the relevant studies in the area of data-driven SV assessment and prioritization (see Figure~\ref{fig:taxonomy}). Specifically, we extracted the themes by grouping related SV assessment or prioritization tasks that the surveyed studies aim to automate/predict using data-driven models. Note that a paper is categorized into more than one theme if that paper develops models for multiple cross-theme tasks.

We acknowledge that there can be other ways to categorize the studies. However, we assert the reliability of our taxonomy as all of our themes (except theme 5) align with the security standards used in practice. For example, Common Vulnerability Scoring System (CVSS)~\cite{cvss} provides a framework to characterize exploitability, impact and severity of SVs (themes 1-3), while Common Weakness Enumeration (CWE)~\cite{cwe} includes many vulnerability types (theme 4). Hence, we believe our taxonomy can help identify and bridge the knowledge gap between the academic literature and industrial practices, making it relevant and potentially beneficial for both researchers and practitioners. Details of each theme in our taxonomy are covered in subsequent sections.

\begin{table}
\fontsize{8}{9}\selectfont
\centering
\caption{List of the surveyed papers in the \textit{Exploit Likelihood} sub-theme of the \textit{Exploitation} theme.
\textbf{Note}: The nature of task of this sub-theme is binary classification of existence/possibility of proof-of-concept and/or real-world exploits.
}
\label{tab:exploit_studies_likelihood}

\begin{tabular}[!t]{|p{1.5cm}|p{6.2cm}|p{5cm}|}

  \hline \textbf{Study} & \textbf{Data source} & \textbf{Data-driven technique}\\\hline

  Bozorgi et al. 2010~\cite{bozorgi2010beyond} & CVE, Open Source Vulnerability Database (OSVDB) & Linear Support Vector Machine (SVM)\\\hline
  Sabottke et al. 2015~\cite{sabottke2015vulnerability} & NVD, Twitter, OSVDB, ExploitDB, Symantec security advisories, private Microsoft security advisories & Linear SVM\\\hline
  Edkrantz et al. 2015~\cite{edkrantz2015predictingthesis,edkrantz2015predicting} & NVD, Recorded Future security advisories, ExploitDB & Na\"ive Bayes, Linear SVM, Random forest\\\hline
  Bullough et al. 2017~\cite{bullough2017predicting} & NVD, Twitter, ExploitDB & Linear SVM\\\hline
      
  Almukaynizi et al.~\cite{almukaynizi2017proactive,almukaynizi2019patch} & NVD, ExploitDB, Zero Day Initiative security advisories \& Darkweb forums/markets & SVM, Random forest, Na\"ive Bayes, Bayesian network, Decision tree, Logistic regression\\\hline
  Xiao et al. 2018~\cite{xiao2018patching} & NVD, SecurityFocus security advisories, Symantec \newline Spam/malicious activities based on daily blacklists from abuseat.org, spamhaus.org, spamcop.net, uceprotect.net, wpbl.info \& list of unpatched SVs in hosts & Identification of malicious activity groups with community detection algorithms + Random forest for exploit prediction\\\hline
  Tavabi et al. 2018~\cite{tavabi2018darkembed} & NVD, 200 sites on Darkweb, ExploitDB, Symantec, Metasploit & Paragraph embedding + Radial basis function kernel SVM\\\hline
  de Sousa et al. 2020~\cite{de2020evaluating} & NVD, Twitter, ExploitDB, Symantec\newline Avast, ESET, Trend Micro security advisories & Linear SVM, Logistic regression, XGBoost, Light Gradient Boosting Machine (LGBM)\\\hline
  Fang et al. 2020~\cite{fang2020fastembed} & NVD, ExploitDB, SecurityFocus, Symantec & fastText + LGBM\\\hline
  Huang et al. 2020~\cite{huang2020poster} & NVD, CVE Details, Twitter, ExploitDB, Symantec security advisories & Random forest\\\hline
  Jacobs et al. 2020~\cite{jacobs2020improving} & NVD, Kenna Security \newline Exploit sources: Exploit DB, Metasploit, FortiGuard Labs, SANS Internet Storm Center, Securewords CTU, Alienvault OSSIM, Canvas/D2 Security's Elliot Exploitation Frameworks, Contagio, Reversing Labs
  & XGBoost\\\hline
  Yin et al. 2020~\cite{yin2020apply} & NVD, ExploitDB, General text: Book Corpus \& Wikipedia for pretraining BERT models & Fine-tuning BERT models pretrained on general text\\\hline
  Bhatt et al. 2021~\cite{bhatt2021exploitability} & NVD, ExploitDB & Features augmented by SV types + Decision tree, Random forest, Na\"ive Bayes, Logistic regression, SVM\\\hline
  Suciu et al. 2021~\cite{suciu2021expected} & NVD, Vulners database, Twitter, Symantec, SecurityFocus, IBM X-Force Threat Intelligence \newline Exploit sources: ExploitDB, Metasploit, Canvas, D2 Security's Elliot, Tenable, Skybox, AlienVault, Contagio & Multi-layer perceptron\\\hline\hline
  
  Younis et al. 2014~\cite{younis2014using} & Vulnerable functions from NVD (Apache HTTP Server project), ExploitDB, OSVDB & SVM\\\hline
  Yan et al. 2017~\cite{yan2017exploitmeter} & Executables (binary code) of 100 Linux applications & Combining ML (Decision tree) output \& fuzzing with a Bayesian network\\\hline
  Tripathi et al. 2017~\cite{tripathi2017exniffer} & Program crashes from VDiscovery~\cite{cha2012unleashing,grieco2016toward} \& LAVA~\cite{dolan2016lava} datasets & Static/Dynamic analysis features + Linear/Radial basis function kernel SVM\\\hline
  Zhang et al. 2018~\cite{zhang2018assisting} & Program crashes from VDiscovery~\cite{cha2012unleashing,grieco2016toward} dataset & $n$-grams of system calls from execution traces + Online passive-aggressive classifier\\\hline
  
\end{tabular}
\end{table}

\section{Exploitation Prediction}
\label{sec:exploit_prediction}

This section covers the \textit{Exploitation} theme that automates the detection and understanding of both Proof-of-Concept (PoC) and real-world exploits\footnote{An \textit{exploit} is a piece of code used to compromise vulnerable software~\cite{sabottke2015vulnerability}. Real-world exploits are harmful \& used in real host/network-based attacks. PoC exploits are unharmful \& used to show the potential threats of SVs in penetration tests.} targeting identified SVs.
This theme outputs the origin of SVs and how/when attackers would take advantage of such SVs to compromise a system of interest, assisting practitioners to quickly react to the more easily exploitable or already exploited SVs.
The papers in this theme can be categorized into three groups/sub-themes: (\textit{i}) Exploit likelihood, (\textit{ii}) Exploit time, (\textit{iii}) Exploit characteristics, as given in Tables~\ref{tab:exploit_studies_likelihood},~\ref{tab:exploit_studies_time} and~\ref{tab:exploit_studies_prop}, respectively.

\subsection{Summary of Primary Studies}
\subsubsection{Exploit Likelihood}
\label{subsubsec:exploit_likelihood}

The first sub-theme is \textit{exploit likelihood} that predicts whether SVs would be exploited in the wild or PoC exploits would be released publicly (see Table~\ref{tab:exploit_studies_likelihood}). In 2010, Bozorgi et al.~\cite{bozorgi2010beyond} were the first to use SV descriptions on Common Vulnerabilities and Exposures (CVE)~\cite{cve} and Open Source Vulnerability Database (OSVDB)\footnote{\url{http://osvdb.org}. Note that this database has been discontinued since 2016.} to predict exploit existence based on the labels on OSVDB. In 2015, Sabottke et al.~\cite{sabottke2015vulnerability} conducted a seminal study that used Linear SVM and SV information on Twitter to predict PoC exploits on ExploitDB~\cite{exploitdb} as well as real-world exploits on OSVDB, Symantec's attack signatures~\cite{symantec} and private Microsoft's security advisories~\cite{ms_security}. These authors urged to explicitly consider real-world exploits as \textit{not} all PoC exploits would result in exploitation in practice.
They also showed SV-related information on Twitter\footnote{\url{https://twitter.com}} can enable earlier detection of exploits than using expert-verified SV sources (e.g., NVD).

Built upon these two foundational studies~\cite{bozorgi2010beyond,sabottke2015vulnerability}, the literature has mainly aimed to improve the performance and applicability of exploit prediction models by leveraging more exploit sources and/or better data-driven techniques/practices.
Many researchers~\cite{edkrantz2015predicting,edkrantz2015predictingthesis,almukaynizi2017proactive,tavabi2018darkembed,almukaynizi2019patch,jacobs2020improving} increased the amount of ground-truth exploits using extensive sources other than ExploitDB and Symantec in~\cite{bozorgi2010beyond,sabottke2015vulnerability}. The sources were security advisories such as Zero Day Initiative~\cite{zeroday_initiative}, Metasploit~\cite{metasploit}, SecurityFocus~\cite{securityfocus}, Recorded Future~\cite{recorded_future}, Kenna Security~\cite{kenna_security}, Avast\footnote{\url{https://avast.com/exploit-protection.php}. This link was provided by de Sousa et al.~\cite{de2020evaluating}, but it is no longer available.}, ESET~\cite{eset}, Trend Micro~\cite{trend_micro}, malicious activities in hosts based on traffic of spam/malicious IP addresses~\cite{xiao2018patching} and Darkweb sites/forums/markets~\cite{nunes2016darknet}. In addition to enriching exploit sources, better data-driven models and practices for exploit prediction were also studied. Ensemble models (e.g., Random forest, eXtreme Gradient Boosting (XGBoost)~\cite{chen2016xgboost}, Light Gradient Boosting Machine (LGBM)~\cite{ke2017lightgbm}) were shown to outperform single-model baselines (e.g., Na\"ive Bayes, SVM, Logistic regression and Decision tree) for exploit prediction~\cite{fang2020fastembed,jacobs2020improving,de2020evaluating,huang2020poster}.
Additionally, Bullough et al.~\cite{bullough2017predicting} identified and addressed several issues with exploit prediction models, e.g., time sensitivity of SV data, already-exploited SVs before disclosure and training data imbalance, helping to improve the practical application of such models. Recently, Yin et al.~\cite{yin2020apply} demonstrated that transfer learning is an alternative solution for improving the performance of exploit prediction with scarcely labeled exploits. Specifically, these authors pre-trained a DL model, BERT~\cite{devlin2018bert}, on massive non-SV sources (e.g., text on Book Corpus~\cite{zhu2015aligning} and Wikipedia~\cite{wikipedia}) and then fine-tuned this pre-trained model on SV data using additional pooling and dense layers. Bhatt et al.~\cite{bhatt2021exploitability} also suggested that incorporating the types of SVs (e.g., SQL injection) into ML models can further enhance the predictive effectiveness. Suciu et al.~\cite{suciu2021expected} empirically showed that unifying SV-related sources used in prior work (e.g., SV databases~\cite{bozorgi2010beyond}, social media~\cite{sabottke2015vulnerability}, SV-related discussions~\cite{tavabi2018darkembed} and PoC code in ExploitDB~\cite{jacobs2020improving}) supports more effective and timely prediction of \textit{functional} exploits~\cite{cvss_v31}.

Besides using SV descriptions as input for exploit prediction, several studies in this sub-theme have also predicted exploits on the code level.
Younis et al.~\cite{younis2014using} predicted the exploitability of vulnerable functions in the Apache HTTP Server project. Specifically, these authors used an SVM model with features extracted from the dangerous system calls~\cite{bernaschi2002remus} in entry points/functions~\cite{manadhata2010attack} and the reachability from any of these entry points to vulnerable functions~\cite{horwitz1990interprocedural}.
Moving from high-level to binary code, Yan et al.~\cite{yan2017exploitmeter} first used a Decision tree to obtain prior beliefs about SV types in 100 Linux applications using static features (e.g., \textit{hexdump}) extracted from executables. Subsequently, they applied various fuzzing tools (i.e., Basic Fuzzing Framework~\cite{basic_fuzzing_framework} and OFuzz~\cite{ofuzz}) to detect SVs with the ML-predicted types. They finally updated the posterior beliefs about the exploitability based on the outputs of the ML model and fuzzers using a Bayesian network. The proposed method outperformed \textit{!exploitable},\footnote{\url{https://microsoft.com/security/blog/2013/06/13/exploitable-crash-analyzer-version-1-6}} a static crash analyzer provided by Microsoft.
Tripathi et al.~\cite{tripathi2017exniffer} also predicted SV exploitability from crashes (i.e., VDiscovery~\cite{cha2012unleashing,grieco2016toward} and LAVA~\cite{dolan2016lava} datasets) using an SVM model and static features from core dumps and dynamic features generated by the Last Branch Record hardware debugging utility.
Zhang et al.~\cite{zhang2018assisting} proposed two improvements to Tripathi et al.~\cite{tripathi2017exniffer}'s approach. These authors first replaced the hardware utility in~\cite{tripathi2017exniffer} that may not be available for resource-constrained devices (e.g., IoT) with sequence/$n$-grams of system calls extracted from execution traces. They also used an online passive-aggressive classifier~\cite{crammer2006online} to enable online/incremental learning of exploitability for new crash batches on-the-fly.

\begin{table}
\fontsize{8}{9}\selectfont
\centering
\caption{List of the surveyed papers in the \textit{Exploit Time} sub-theme of the \textit{Exploitation} theme.}
\label{tab:exploit_studies_time}

\begin{tabular}[!t]{|p{1.5cm}|p{4cm}|p{3.5cm}|p{3cm}|}

  \hline \textbf{Study} & \textbf{Nature of task} & \textbf{Data source} & \textbf{Data-driven technique}\\\hline

  Bozorgi et al. 2010~\cite{bozorgi2010beyond} & \textit{Binary classification}: Likelihood that SVs would be exploited within 2 to 30 days after disclosure & CVE, OSVDB & Linear SVM\\\hline
  
  Edkrantz 2015~\cite{edkrantz2015predictingthesis} & \multirowcell{2}[0ex][l]{\textit{Binary classification}: Likelihood of\\SV exploits within 12 months after\\disclosure} & NVD, ExploitDB, Recorded Future security advisories & SVM, K-Nearest Neighbors (KNN), Na\"ive Bayes, Random forest\\\hhline{-~*{2}{-}}
  Jacobs et al. 2019~\cite{jacobs2019exploit} & & NVD, Kenna Security \newline Exploit sources: Exploit DB, Metasploit, D2 Security's Elliot \& Canvas Exploitation Frameworks, Fortinet, Proofpoint, AlienVault \& GreyNoise & Logistic regression\\\hline
  
  Chen et al. 2019~\cite{chen2019using,chen2019vest} & \textit{Binary classification}: Likelihood that SVs would be exploited within 1/3/6/9/12 months after disclosure \newline \textit{Regression}: number of days until SV exploits after disclosure & CVE, Twitter, ExploitDB, Symantec security advisories & Graph neural network embedding + Linear regression, Bayes, Random forest, XGBoost, Lasso/Ridge regression\\\hline

\end{tabular}
\end{table}

\subsubsection{Exploit Time}
\label{subsubsec:exploit_time}

After predicting the likelihood of SV exploits in the previous sub-theme, this sub-theme provides more fine-grained information about \textit{exploit time} (see Table~\ref{tab:exploit_studies_time}).
Besides performing binary classification of exploits, Bozorgi et al.~\cite{bozorgi2010beyond} and Edkrantz~\cite{edkrantz2015predictingthesis} also predicted the time frame (2-30 days in~\cite{bozorgi2010beyond} and 12 months in~\cite{edkrantz2015predictingthesis}) within which exploits would happen after the disclosure of SVs. Jacobs et al.~\cite{jacobs2019exploit} then leveraged multiple sources containing both PoC and real-world exploits, as given in Table~\ref{tab:exploit_studies_time}, to improve the number of labeled exploits, enhancing the prediction of exploit appearance within 12 months. Chen et al.~\cite{chen2019using} predicted whether SVs would be exploited within 1-12 months and the exploit time (number of days) after SV disclosure using Twitter data. The authors proposed a novel regression model whose feature embedding was a multi-layer graph neural network~\cite{kivela2014multilayer} capturing the content and relationships among tweets, respective tweets' authors and SVs. The proposed model outperformed many baselines and was integrated into the VEST system~\cite{chen2019vest} to provide timely SV assessment information for practitioners. To the best of our knowledge, at the time of writing, Chen et al.~\cite{chen2019using,chen2019vest} have been the only ones pinpointing the exact exploit time of SVs rather than large/uncertain time-frames (e.g., months) in other studies, helping practitioners to devise much more fine-grained remediation plans.

\subsubsection{Exploit Characteristics}\label{subsubsec:exploit_prop}

\textit{Exploit characteristics} is the final sub-theme that reveals various requirements/means of exploits (see Table~\ref{tab:exploit_studies_prop}), informing the potential scale of SVs; e.g., remote exploits likely affect more systems than local ones.
The commonly used outputs are the Exploitability metrics provided by versions 2~\cite{cvss_v2} and 3~\cite{cvss_v3,cvss_v31} of Common Vulnerability Scoring System (CVSS).

\begin{table}
\fontsize{8}{9}\selectfont
\centering
\caption{List of the surveyed papers in the \textit{Exploit Characteristics} sub-theme of the \textit{Exploitation} theme.}
\label{tab:exploit_studies_prop}

\begin{tabular}[!t]{|p{1.5cm}|p{4.5cm}|p{3.2cm}|p{3cm}|}

  \hline \textbf{Study} & \textbf{Nature of task} & \textbf{Data source} & \textbf{Data-driven technique}\\\hline

  Yamamoto et al. 2015~\cite{yamamoto2015text} & \multirowcell{3}[0ex][l]{\textit{Multi-class classification}: CVSS v2\\(Access Vector \& Access Complexity\\ metrics)\\\\ \textit{Binary classification}: CVSS v2\\(Authentication metric)} & NVD & Supervised Latent Dirichlet Allocation (LDA)\\\hhline{-~*{2}{-}}
  Wen et al. 2015~\cite{wen2015novel} & & NVD, OSVDB, SecurityFocus, IBM X-Force & Radial basis function kernel SVM\\\hhline{-~*{2}{-}}
  Le et al. 2019~\cite{le2019automated} & & NVD & Concept-drift-aware models with Na\"ive Bayes, KNN, Linear SVM, Random forest, XGBoost, LGBM\\\hline
 
  Toloudis et al. 2016~\cite{toloudis2016associating} & \textit{Correlation analysis}: CVSS v2 & NVD & Principal component analysis \& Spearman correlation coefficient\\\hline
 
  Ognawala et al. 2018~\cite{ognawala2018automatically} & \multirowcell{2}[0ex][l]{\textit{Multi-class classification}: CVSS v3\\(Attack Vector, Attack Complexity \&\\ Privileges Required metrics)\\\\ \textit{Binary classification}: CVSS v3 (User\\ Interaction metric)} & NVD (buffer overflow SVs) \& Source code of vulnerable software/components & Combining static analysis tool (Macke~\cite{ognawala2016macke}) \& ML classifiers (Na\"ive Bayes \& Random forest)\\\hhline{-~*{2}{-}}
  Chen et al. 2019~\cite{chen2019vest} & & CVE, NVD, Twitter & Graph convolutional network\\\hline
 
  Elbaz et al. 2020~\cite{elbaz2020fighting} & \textit{Multi-class/Binary classification}: CVSS v2/v3 & NVD & Mapping outputs of Linear regression to CVSS metrics with closest values\\\hhline{-~*{2}{-}}
  Jiang et al. 2020~\cite{jiang2020approach} & & NVD, ICS Cert, Vendor websites (Resolve inconsistencies with a majority vote) & Logistic regression\\\hline
 
  Gawron et al. 2017~\cite{gawron2017automatic} & \textit{Multi-target classification}: CVSS v2 & NVD & Na\"ive Bayes, Multi-layer Perceptron (MLP)\\\hhline{-~*{2}{-}}
  Spanos et al. 2018~\cite{spanos2018multi} & & NVD & Random forest, boosting model, Decision tree\\\hline
 
  Gong et al. 2019~\cite{gong2019joint} & \textit{Multi-task classification}: CVSS v2 & NVD & Bi-LSTM with attention mechanism\\\hline\hline
 
  Chen et al. 2010~\cite{chen2010categorization} & \textit{Multi-class classification}: Platform-specific vulnerability locations (Local, Remote, Local area network) \& vulnerability causes (e.g., Access/Input/Origin validation error) & NVD, Secunia vulnerability database, SecurityFocus, IBM X-Force & Linear SVM\\\hline

  Ruohonen et al. 2017~\cite{ruohonen2017classifying} & \textit{Binary classification}: Web-related exploits or not & ExploitDB & LDA + Random forest\\\hline
 
  Aksu et al. 2018~\cite{aksu2018automated} & \textit{Multi-class classification}: author-defined pre-/post-condition privileges (None, OS (Admin/User), App (Admin/User)) & NVD & RBF network, Linear SVM, NEAT~\cite{stanley2002evolving}, MLP\\\hhline{-~*{2}{-}}
  Liu et al. 2019~\cite{liu2019automated} & & NVD & Information gain + Convolutional neural network\\\hline
  
  Kanakogi et al. 2021~\cite{kanakogi2021tracing} & \textit{Multi-class classification}: Common Attack Pattern Enumeration and Classification (CAPEC) & NVD, CAPEC & Doc2vec/tf-idf with cosine similarity\\\hline
  
\end{tabular}
\end{table}

Many studies have focused on predicting and analyzing version 2 of CVSS exploitability metrics (i.e., Access Vector, Access Complexity and Authentication). Yamamoto et al.~\cite{yamamoto2015text} were the first one to leverage descriptions of SVs on NVD together with a supervised Latent Dirichlet Allocation topic model~\cite{blei2007supervised} to predict these CVSS metrics.
Subsequently, Wen et al.~\cite{wen2015novel} used  Radial Basis Function (RBF)-kernel SVM and various SV databases/advisories other than NVD (e.g., SecurityFocus, OSVDB and IBM X-Force~\cite{xforce}) to predict the metrics.
Le et al.~\cite{le2019automated} later showed that the prediction of CVSS metrics suffered from the \textit{concept drift} issue; i.e., descriptions of new SVs may contain Out-of-Vocabulary terms for prediction models.
They proposed to combine sub-word features with traditional Bag-of-Word (BoW) features to infer the semantics of novel terms/words from existing ones, helping assessment models be more robust against concept drift. Besides prediction, Toloudis et al.~\cite{toloudis2016associating} used principal component analysis~\cite{wold1987principal} and Spearman's $\rho$ correlation coefficient to reveal the predictive contribution of each word in SV descriptions to each CVSS metric. However, this technique does not directly produce the value of each metric.

Recently, several studies have started to predict CVSS version 3 exploitability metrics including the new Privileges and User Interactions.
Ognawala et al.~\cite{ognawala2018automatically} fed the features generated by a static analysis tool, Macke~\cite{ognawala2016macke}, to a Random forest model to predict these CVSS version 3 metrics for vulnerable software/components.
Later, Chen et al.~\cite{chen2019vest} found that many SVs were disclosed on Twitter before on NVD. Therefore, these authors developed a system built on top of a Graph Convolutional Network~\cite{kipf2016semi} capturing the content and relationships of related Twitter posts about SVs to enable more timely prediction of the CVSS version 3 metrics.
Elbaz et al.~\cite{elbaz2020fighting} developed a linear regression model to predict the numerical output of each metric and then obtained the respective categorical value with the numerical value closest to the predicted value. For example, a predicted value of 0.8 for Attack Vector CVSS v3 is mapped to \textit{Network} (0.85)~\cite{cvss_v3}.
To prepare a clean dataset to predict these CVSS metrics, Jiang et al.~\cite{jiang2020approach} replaced inconsistent CVSS values in various SV sources (i.e., NVD, ICS CERT and vendor websites) with the most frequent value.

Instead of building a separate model for each CVSS metric, there has been another family of approaches predicting these metrics using a single model to increase efficiency. Gawron et al.~\cite{gawron2017automatic} and Spanos et al.~\cite{spanos2018multi} predicted multiple CVSS metrics as a unique string instead of individual values. The output of each metric is then extracted from the concatenated string.
Later, Gong et al.~\cite{gong2019joint} adopted the idea of a unified model from the DL perspective by using the multi-task learning paradigm~\cite{zhang2021survey} to predict CVSS metrics simultaneously. The model has a feature extraction module (based on a Bi-LSTM model with attention mechanism~\cite{bahdanau2014neural}) shared among all the CVSS metrics/tasks, yet specific prediction head/layer for each metric/task. This model outperformed single-task counterparts while requiring much less time to (re-)train.

Although CVSS exploitability metrics were most commonly used, several studies used other schemes for characterizing exploitation. Chen et al.~\cite{chen2010categorization} used Linear SVM and SV descriptions to predict multiple SV characteristics, including three \textit{SV locations} (i.e., Local, LAN and Remote) on SecurityFocus~\cite{securityfocus} and Secunia~\cite{secunia} databases as well as 11 \textit{SV causes}\footnote{Access/Input/Origin validation error, Atomicity/Configuration/Design/Environment/Serialization error, Boundary condition error, Failure on exceptions, Race condition error} on SecurityFocus.
Regarding the exploit types, Rouhonen et al.~\cite{ruohonen2017classifying} used LDA~\cite{blei2003latent} and Random forest to classify whether an exploit would affect a web application. This study can help find relevant exploits in components/sub-systems of a large system.
For privileges, Aksu et al.~\cite{aksu2018automated} extended the Privileges Required metric of CVSS by incorporating the context (i.e., Operating system or Application) to which privileges are applied (see Table~\ref{tab:exploit_studies_prop}).
They found MLP~\cite{hastie2009elements} to be the best-performing model for obtaining these privileges from SV descriptions. They also utilized the predicted privileges to generate attack graphs (sequence of attacks from source to sink nodes).
Liu et al.~\cite{liu2019automated} advanced this task by combining information gain for feature selection and Convolutional Neural Network (CNN)~\cite{kim2014convolutional} for feature extraction.
Regarding attack patterns, Kanakogi et al.~\cite{kanakogi2021tracing} found Doc2vec~\cite{le2014distributed} to be more effective than term-frequency inverse document frequency (tf-idf) when combined with cosine similarity to find the most relevant Common Attack Pattern Enumeration and Classification (CAPEC)~\cite{capec} for a given SV on NVD. Such attack patterns can manifest how identified SVs can be exploited by adversaries, assisting the selection of suitable countermeasures.

\subsection{Theme Discussion}
In the \textit{Exploitation} theme, the primary tasks are binary classification of whether Proof-of-Concept (PoC)/real-world exploits of SVs would appear and multi-classification of exploit characteristics based on CVSS. PoC exploits mostly come from ExploitDB~\mbox{\cite{exploitdb}}; whereas, real-world exploits, despite coming from multiple sources, are still much scarcer than PoC counterparts. Consequently, the models predicting real-world exploits have generally performed worse than those for PoC exploits. Similarly, the performance of the models determining CVSS v3 exploitability metrics has been mostly lower than that of the CVSS v2 based models. However, real exploits and CVSS v3 are usually of more interest to the community. The former can lead to real cyber-attacks and the latter is the current standard in practice. To improve the performance of these practical tasks, future work can collect more exploit-related data from the relevant yet under-explored sources (see section~\mbox{\ref{subsec:data_avail}}), as well as adapt the patterns learned from PoC exploits and old CVSS versions to real exploits and newer CVSS versions, respectively, e.g., using transfer learning~\mbox{\cite{pan2009survey}}.

There are other under-explored tasks targeting fine-grained prediction of exploits.
Mitigation of exploits in practice usually requires more information besides simply determining whether an SV would be exploited.
Information gathered from predicting \textit{when} and \textit{how} the exploits would happen is also needed to devise better SV fixing prioritization and mitigation plans. VEST~\mbox{\cite{chen2019vest}} is one of the first and few systems aiming to provide such all-in-one information about SV exploitation. However, this system currently only uses data from NVD/CVE and Twitter, which can be extended to incorporate more (exploit) sources and more sophisticated data-driven techniques in the future.

Most of the current studies have used SV descriptions on NVD and other security advisories to predict the exploitation-related metrics. This is surprising as SV descriptions do not contain root causes of SVs. Instead, SVs are rooted in source code, yet there is little work on code-based exploit prediction. So far, Younis et al.~\mbox{\cite{younis2014using}} have been among the few ones using source code for exploit prediction, but their approach still requires manual identification of dangerous function calls in C/C++. More work is required to employ data-driven approaches to alleviate the need for manually defined rules to improve the effectiveness and generalizability of code-based exploit prediction.

\vspace{-1pt}
\section{Impact Prediction}
\label{sec:impact_prediction}

This section describes the \textit{Impact} theme that determines the (negative) effects that SVs have on a system of interest if such SVs are exploited. There are five key tasks that the papers in this theme have automated/predicted: (\textit{i}) Confidentiality impact, (\textit{ii}) Integrity impact, (\textit{iii}) Availability impact, (\textit{iv}) Scope and (\textit{v}) Custom vulnerability consequences (see Table~\ref{tab:impact_studies}). 

\begin{table}[!t]
\fontsize{8}{9}\selectfont
\caption{List of the surveyed papers in the \textit{Impact} theme. \textbf{Note}: We grouped the first four sub-themes as they were mostly predicted together.}
\label{tab:impact_studies}
\centering
\begin{tabular}{|p{1.5cm}|p{3.8cm}|p{3.2cm}|p{3.8cm}|}

 \hline \multicolumn{1}{|c|}{\textbf{Study}} & \multicolumn{1}{c|}{\textbf{Nature of task}} & \multicolumn{1}{c|}{\textbf{Data source}} & \multicolumn{1}{c|}{\textbf{Data-driven technique}}\\\hline
 
 \multicolumn{4}{|c|}{\cellcolor[HTML]{C0C0C0}}\\
 \multicolumn{4}{|c|}{\multirow{-2}{*}{\cellcolor[HTML]{C0C0C0} \textbf{Sub-themes: 1. Confidentiality, 2. Integrity, 3. Availability \& 4. Scope (only in CVSS v3)}}}\\\hline

 Yamamoto et al. 2015~\cite{yamamoto2015text} & \textit{Multi-class classification}: CVSS v2 & NVD & Supervised Latent Dirichlet Allocation\\\hhline{-~*{2}{-}}
 Wen et al. 2015~\cite{wen2015novel} & & NVD, OSVDB, SecurityFocus, IBM X-Force & Radial basis function kernel SVM\\\hhline{-~*{2}{-}}
 Le et al. 2019~\cite{le2019automated} & & NVD & Concept-drift-aware models with Na\"ive Bayes, KNN, Linear SVM, Random forest, XGBoost, LGBM\\\hline
 
  Toloudis et al. 2016~\cite{toloudis2016associating} & \textit{Correlation analysis}: CVSS v2 & NVD & Principal component analysis \& Spearman correlation coefficient\\\hline
 
  Ognawala et al. 2018~\cite{ognawala2018automatically} & \multirowcell{2}[0ex][l]{\textit{Multi-class classification}: CVSS v3\\\\ \textit{Binary classification}: Scope in\\ CVSS v3} & NVD (buffer overflow SVs) \& Source code of vulnerable software/components & Combining static analysis tool (Macke~\cite{ognawala2016macke}) \& ML classifiers (Na\"ive Bayes \& Random forest)\\\hhline{-~*{2}{-}}
  Chen et al. 2019~\cite{chen2019vest} & & CVE, NVD, Twitter & Graph convolutional network\\\hline
 
 Elbaz et al. 2020~\cite{elbaz2020fighting} & \multirowcell{2}[0ex][l]{\textit{Multi-class classification}: CVSS\\ v2/v3\\\\ \textit{Binary classification}: Scope in\\ CVSS v3} & NVD & Mapping outputs of Linear regression outputs to CVSS metrics with closest values\\\hhline{-~*{2}{-}}
 Jiang et al. 2020~\cite{jiang2020approach} & & NVD, ICS Cert, Vendor websites (Resolve inconsistencies with a majority vote) & Logistic regression\\\hline
 
 Gawron et al. 2017~\cite{gawron2017automatic} & \textit{Multi-target classification}: CVSS v2 & NVD & Na\"ive Bayes, MLP\\\hhline{-~*{2}{-}}
 Spanos et al. 2018~\cite{spanos2018multi} & & NVD & Random forest, boosting model, Decision tree\\\hline
 
 Gong et al. 2019~\cite{gong2019joint} & \textit{Multi-task classification}: CVSS v2 & NVD & Bi-LSTM with attention mechanism\\\hline
 
 \multicolumn{4}{|c|}{\cellcolor[HTML]{C0C0C0}}\\
 \multicolumn{4}{|c|}{\multirow{-2}{*}{\cellcolor[HTML]{C0C0C0} \textbf{Sub-theme: 5. Custom Vulnerability Consequences}}}\\\arrayrulecolor{black}\hline
 
 Chen et al. 2010~\cite{chen2010categorization} & \textit{Multi-label classification}: Platform-specific impacts (e.g., Gain system access) & NVD, Secunia vulnerability database, SecurityFocus, IBM X-Force & Linear SVM\\\hline

\end{tabular}
\vspace{-3pt}
\end{table}

\subsection{Summary of Primary Studies}
\subsubsection{Confidentiality, Integrity, Availability, and Scope}\label{subsubsec:cvss_impact}

A majority of the papers have focused on the impact metrics provided by CVSS, including versions 2~\cite{cvss_v2} and 3~\cite{cvss_v3,cvss_v31}.
Versions 2 and 3 share three impact metrics \textit{Confidentiality}, \textit{Integrity} and \textit{Availability}. Version 3 also has a new metric, \textit{Scope}, that specifies whether an exploited SV would affect only the system that contains the SV. For example, \textit{Scope} changes when an SV occurring in a virtual machine affects the whole host machine, in turn increasing individual impacts.

The studies that predicted the CVSS impact metrics are mostly the same as the ones predicting the CVSS exploitability metrics in section~\ref{sec:exploit_prediction}.
Given the overlap, we hereby only describe the main directions and techniques of the \textit{Impact}-related tasks rather than iterating the details of each study.
Overall, a majority of the work has focused on classifying CVSS impact metrics (versions 2 and 3) using three main learning paradigms: single-task~\cite{yamamoto2015text,wen2015novel,le2019automated,ognawala2018automatically,chen2019vest,elbaz2020fighting,jiang2020approach}, multi-target~\cite{gawron2017automatic,spanos2018multi} and multi-task~\cite{gong2019joint} learning. Instead of developing a separate prediction model for each metric like the single-task approach, multi-target and multi-task approaches only need a single model for all tasks. Multi-target learning predicts concatenated output; whereas, multi-task learning uses shared feature extraction for all tasks and task-specific softmax layers to determine the output of each task. These three learning paradigms were powered by applying and/or customizing a wide range of data-driven methods. The first method was to use single ML classifiers like supervised Latent Dirichlet Allocation~\cite{yamamoto2015text}, Principal component analysis~\cite{toloudis2016associating}, Na\"ive Bayes~\cite{le2019automated,ognawala2018automatically,gawron2017automatic}, Logistic regression~\cite{jiang2020approach}, Kernel-based SVM~\cite{wen2015novel}, Linear SVM~\cite{le2019automated}, KNN~\cite{le2019automated} and Decision tree~\cite{spanos2018multi}. Other studies employed ensemble models combining the strength of multiple single models such as Random forest~\cite{le2019automated,ognawala2018automatically}, boosting model~\cite{spanos2018multi} and XGBoost/LGBM~\cite{le2019automated}. Recently, more studies moved towards more sophisticated DL architectures such as MLP~\cite{gawron2017automatic}, attention-based (Bi-)LSTM~\cite{gong2019joint} and graph neural network~\cite{chen2019vest}. Ensemble and DL models usually beat the single ones, but there is a lack of direct comparisons between these two emerging model types.

\subsubsection{Custom Vulnerability Consequences}\label{subsubsec:other_consequences}

To devise effective remediation strategies for a system of interest in practice, practitioners may want to know \textit{custom vulnerability consequences} which are more interpretable than the levels of impact provided by CVSS. Chen et al.~\cite{chen2010categorization} curated a list of 11 vulnerability consequences\footnote{Gain system access, Bypass security, Configuration manipulation, Data/file manipulation, Denial of Service, Privilege escalation, Information leakage, Session hijacking, Cross-site scripting (XSS), Source spoofing, Brute-force proneness.} from X-Force~\cite{xforce} and Secunia~\cite{secunia} vulnerability databases.
They then used a Linear SVM model to perform multi-label classification of these consequences for SVs, meaning that an SV can lead to more than one consequence. To the best of our knowledge, this is the only study that has pursued this research direction so far.

\subsection{Theme Discussion}
In the \textit{Impact} theme, the common task is to predict the impact base metrics provided by CVSS versions 2 and 3. Similar to the Exploitation theme, the models for CVSS v3 still require more attention and effort from the community to reach the same performance level as the models for CVSS v2. These impact metrics are also usually predicted together with the exploitability metrics given their similar nature (multi-class classification) using either task-wise models or a unified (multi-target or multi-task) model. Multi-target and multi-task learning are promising as they can reduce the time for continuous (re)training and maintenance when deployed in production.

Besides CVSS impact metrics, other fine-grained SV consequences have also been explored~\mbox{\cite{chen2010categorization}}, but there is still no widely accepted taxonomy for such consequences. Thus, these consequences have seen less adoption in practice than CVSS metrics, despite being potentially useful by providing more concrete information about what assets/components in a system that an SV can compromise. In section~\mbox{\ref{subsubsec:timely_finegrained_prediction}}, we suggest potential ways to create a systematic taxonomy of custom SV consequences  to pave the way for more data-driven research in this direction.

\section{Severity Prediction}
\label{sec:severity_prediction}

This section discusses the work in the \textit{Severity} theme.
Severity is often a function/combination of Exploitation (section~\ref{sec:exploit_prediction}) and Impact (section~\ref{sec:impact_prediction}). SVs with higher severity usually require more urgent remediation.
There are three main prediction tasks in this theme: (\textit{i}) Severe vs. Non-severe, (\textit{ii}) Severity levels and (\textit{iii}) Severity score, shown in Tables~\ref{tab:severity_binary_studies},~\ref{tab:severity_level_studies} and~\ref{tab:severity_score_studies}, respectively.

Similar to the \textit{Exploitation} and \textit{Impact} themes, many studies in the \textit{Severity} theme have used CVSS versions 2 and 3. According to both CVSS versions, the severity score shares the same range from 0 to 10, with an increment of 0.1.
Based on the score, the existing studies have either defined a threshold to decide whether an SV is severe (requiring high attention), or predicted levels/groups of severity score that require a similar amount of attention or determined the raw score value.

\subsection{Summary of Primary Studies}
\subsubsection{Severe vs. Non-Severe}\label{subsubsec:severe_or_not}

\begin{table}
\fontsize{8}{9}\selectfont
\centering
\caption{List of the surveyed papers in the \textit{Severe vs. Non-Severe} sub-theme of the \textit{Severity} theme.
\textbf{Note}: The nature of task here is binary classification of severe SVs with High/Critical CVSS v2/v3 severity levels.
}
\label{tab:severity_binary_studies}

\begin{tabular}{|p{1.5cm}|p{6cm}|p{5.3cm}|}
 \hline \textbf{Study} & \textbf{Data source (software project)} & \textbf{Data-driven technique}\\\hline

 Kudjo et al. 2019~\cite{kudjo2019improving} & NVD (Mozilla Firefox, Google Chrome, Internet Explorer, Microsoft Edge, Sea Monkey, Linux Kernel, Windows 7, Windows 10, Mac OS, Chrome OS) & Term frequency \& inverse gravity moment weighting + KNN, Decision tree, Random forest\\\hline
 Chen et al. 2020~\cite{chen2020automatic} & NVD (Adobe Flash Player, Enterprise Linux, Linux Kernel, Foxit Reader, Safari, Windows 10, Microsoft Office, Oracle Business Suites, Chrome, QuickTime) & Term frequency \& inverse gravity moment weighting + KNN, Decision tree, Na\"ive Bayes, SVM, Random forest\\\hline
 Kudjo et al. 2020~\cite{kudjo2020effect} & NVD (Google Chrome, Mozilla Firefox, Internet Explorer and Linux Kernel) & Find the best smallest training dataset using KNN, Logistic regression, MLP, Random forest\\\hline
 Malhotra et al. 2021~\cite{malhotra2021severity} & NVD (Apache Tomcat) & Chi-square/Information gain + bagging technique, Random forest, Na\"ive Bayes, SVM\\\hline
     
\end{tabular}
\end{table}

The first group of studies have classified whether an SV is \textit{severe or non-severe}, making it a binary classification problem (see Table~\ref{tab:severity_binary_studies}). These studies have typically selected severe SVs as the ones with at least High severity level (i.e., CVSS severity score $\geq$ 7.0). Kudjo et al.~\cite{kudjo2019improving} showed that using term frequency (BoW) with inverse gravity moment weighting~\cite{chen2016turning} to extract features from SV descriptions can enhance the performance of ML models (i.e., KNN, Decision tree and Random forest) in predicting the severity of SVs. Later, Chen et al.~\cite{chen2020automatic} confirmed that this feature extraction method was also effective for more projects and classifiers (e.g., Na\"ive Bayes and SVM). Besides investigating feature extraction, Kudjo et al.~\cite{kudjo2020effect} also highlighted the possibility of finding Bellwether, i.e., the smallest set of data that can be used to train an optimal prediction model, for classifying severity. Recently, Malhotra et al.~\cite{malhotra2021severity} revisited this task by showing that Chi-square and information gain can be effective dimensionality reduction techniques for multiple classifiers, i.e., bagging technique, Random forest, Na\"ive Bayes and SVM.

\subsubsection{Severity Levels}\label{subsec:severity_levels}

\begin{table}
\fontsize{8}{9}\selectfont
\centering
\caption{List of the surveyed papers in the \textit{Severity Levels} sub-theme of the \textit{Severity} theme.}
\label{tab:severity_level_studies}

\begin{tabular}{|p{1.5cm}|p{4cm}|p{3cm}|p{3.8cm}|}
 \hline  \textbf{Study} & \textbf{Nature of task} & \textbf{Data source} & \textbf{Data-driven technique}\\\hline
 
 Spanos et al. 2017~\cite{spanos2017assessment} & \textit{Multi-class classification}: NVD severity levels based on CVSS v2 \& WIVSS (High, Medium, Low) & NVD & Decision tree, SVM, MLP\\\hline

 Wang et al. 2019~\cite{wang2019intelligent} & \multirowcell{2}[0ex][l]{\textit{Multi-class classification}: NVD\\severity levels based on CVSS v2\\(High, Medium, Low)} & NVD (XSS attacks) & XGBoost, Logistic regression, SVM, Random forest\\\hhline{-~*{2}{-}}
 Le et al. 2019~\cite{le2019automated} & & NVD & Concept-drift-aware models with Na\"ive Bayes, KNN, Linear SVM, Random forest, XGBoost, LGBM\\\hhline{-~*{2}{-}}
 Liu et al. 2019~\cite{liu2019vulnerability} & & NVD, China National Vulnerability Database (XSS attacks) & Recurrent Convolutional Neural Network (RCNN), Convolutional Neural Network (CNN), Long-Short Term Memory (LSTM)\\\hhline{-~*{2}{-}}
 Sharma et al. 2020~\cite{sharma2021software} & & CVE Details & CNN\\\hline

 Han et al. 2017~\cite{han2017learning} & \multirowcell{2}[0ex][l]{\textit{Multi-class classification}: Atlassian\\categories of CVSS severity score\\(Critical, High, Medium, Low)} & CVE Details & 1-layer CNN, 2-layer CNN, CNN-LSTM, Linear SVM\\\hhline{-~*{2}{-}}
 Sahin et al. 2019~\cite{sahin2019conceptual} & & NVD & 1-layer CNN, LSTM, XGBoost, Linear SVM\\\hhline{-~*{2}{-}}
 Nakagawa et al. 2019~\cite{nakagawa2019character} & & CVE Details & Character-level CNN vs. Word-based CNN + Linear SVM\\\hline
 
 Gong et al. 2019~\cite{gong2019joint} & \textit{Multi-task classification}: Atlassian categories of CVSS severity score (Critical, High, Medium, Low) & CVE Details & Bi-LSTM with attention mechanism\\\hline
 
 Chen et al. 2010~\cite{chen2010categorization} & \textit{Multi-class classification}: severity levels of Secunia (Extremely/highly/ moderately/less/non- critical) & CVE, Secunia vulnerability database, SecurityFocus, IBM X-Force & Linear SVM\\\hline
 
 Zhang et al. 2020~\cite{zhang2020general} & \textit{Multi-class classification}: Platform-specific levels (High/Medium/Low) & China National Vulnerability Database & Logistic regression, Linear discriminant analysis, KNN, CART, SVM, bagging/boosting models\\\hline

 Khazaei et al. 2016~\cite{khazaei2016automatic} & \textit{Multi-class classification}: 10 severity score bins (one unit/bin) & CVE \& OSVDB & Linear SVM, Random forest, Fuzzy system\\\hline

\end{tabular}
\end{table}

Rather than just performing binary classification of whether an SV is severe, several studies have identified one among multiple \textit{severity levels} that an SV belongs to (see Table~\ref{tab:severity_level_studies}). This setting can be considered as multi-class classification. Spanos et al.~\cite{spanos2017assessment} were to first one to show the applicability of ML to classify SVs into one of the three severity levels using SV descriptions. These three levels are provided by NVD and based on the severity score of CVSS version 2~\cite{cvss_v2} and WIVSS~\cite{spanos2013wivss}, i.e., Low (0.0 -- 3.9), Medium (4.0 -- 6.9), High (7.0 -- 10.0). Note that WIVSS assigns different weights for the Confidentiality, Integrity and Availability impact metrics of CVSS, enhancing the ability to capture varied contributions of these impacts to the final severity score. Later, Wang et al.~\cite{wang2019intelligent} showed that XGBoost~\cite{chen2016xgboost} performed the best among the investigated ML classifiers for predicting these three NVD-based severity levels. Le et al.~\cite{le2019automated} also confirmed that ensemble methods (e.g., XGBoost~\cite{chen2016xgboost}, LGBM~\cite{ke2017lightgbm} and Random forest) outperformed single models (e.g., Na\"ive Bayes, KNN and SVM) for this task. Predicting severity levels has also been tackled with DL techniques~\cite{liu2019vulnerability,sharma2021software} such as Recurrent Convolutional Neural Network (RCNN)~\cite{lai2015recurrent}, Convolutional Neural Network (CNN)~\cite{kim2014convolutional}, Long-Short Term Memory (LSTM)~\cite{hochreiter1997long}. These studies showed potential performance gain of DL models compared to traditional ML counterparts. Han et al.~\cite{han2017learning} showed that DL techniques (i.e., 1-layer CNN) also achieved promising results for predicting a different severity categorization, namely Atlassian's levels.\footnote{\url{https://www.atlassian.com/trust/security/security-severity-levels}} Such findings were successfully replicated by Sahin et al.~\cite{sahin2019conceptual}. Nakagawa et al.~\cite{nakagawa2019character} further enhanced the DL model performance for the same task by incorporating the character-level features into a CNN model~\cite{zhang2015character}. Complementary to performance enhancement, Gong et al.~\cite{gong2019joint} proposed to predict these severity levels concurrently with other CVSS metrics in a single model using multi-task learning~\cite{zhang2021survey} powered by an attention-based Bi-LSTM shared feature extraction model. The unified model was demonstrated to increase both the prediction effectiveness and efficiency.
Besides Atlassian's categories, several studies applied ML models to predict severity levels on other platforms such as Secunia~\cite{chen2010categorization} and China National Vulnerability Database\footnote{\url{https://www.cnvd.org.cn}}~\cite{zhang2020general}. Instead of using textual categories, Khazaei et al.~\cite{khazaei2016automatic} divided the CVSS severity score into 10 bins with 10 increments each (e.g., values of 0 -- 0.9 are in one bin) and obtained decent results (86-88\% Accuracy) using Linear SVM, Random forest and Fuzzy system.

\subsubsection{Severity Score}\label{subsec:severity_score}

\begin{table}
\fontsize{8}{9}\selectfont
\centering
\caption{List of the surveyed papers in the \textit{Severity Score} sub-theme of the \textit{Severity} theme. \textbf{Notes}: \textsuperscript{\textdagger}denotes that the severity score is computed from ML-predicted base metrics using the formula provided by an assessment framework (CVSS and/or WIVSS).}
\label{tab:severity_score_studies}

\begin{tabular}{|p{1.5cm}|p{3cm}|p{3.2cm}|p{4.5cm}|}
 \hline \textbf{Study} & \textbf{Nature of task} & \textbf{Data source} & \textbf{Data-driven technique}\\\hline

 Sahin et al. 2019~\cite{sahin2019conceptual} & \textit{Regression}: CVSS v2 (0-10) & NVD & 1-layer CNN, LSTM, XGBoost regressor, Linear regression\\\hhline{-~*{2}{-}}
 Wen et al. 2015~\cite{wen2015novel} & & OSVDB, SecurityFocus, IBM X-Force & Radial basis function kernel SVM\textsuperscript{\textdagger}\\\hline
 
 Ognawala et al. 2018~\cite{ognawala2018automatically} & \textit{Regression}: CVSS v3 (0-10) & NVD (buffer overflow SVs) & Combining a static analysis tool (Macke~\cite{ognawala2016macke}) \& ML classifiers (Na\"ive Bayes \& Random forest)\textsuperscript{\textdagger}\\\hhline{-~*{2}{-}}
 Chen et al. 2019~\cite{chen2019vest,chen2019vase} & & CVE, NVD, Twitter & Graph convolutional network\\\hhline{-~*{2}{-}}
 Anwar et al. 2020~\cite{anwar2020cleaning} & & NVD & Linear regression, Support vector regression, CNN, MLP\\\hline
 
 Elbaz et al. 2020~\cite{elbaz2020fighting} & \textit{Regression}: CVSS v2/v3 (0-10) & NVD & Mapping outputs of Linear regression to CVSS metrics with closest values\textsuperscript{\textdagger}\\\hhline{-~*{2}{-}}
 Jiang et al. 2020~\cite{jiang2020approach} & & NVD, ICS Cert, Vendor websites (Resolve inconsistencies with a majority vote) & Logistic regression\textsuperscript{\textdagger}\\\hline
 
 Spanos et al. 2018~\cite{spanos2018multi} & \textit{Regression}: CVSS v2 \& WIVSS (0-10) & NVD & Random forest, boosting model, Decision tree\textsuperscript{\textdagger}\\\hline
 
 Toloudis et al. 2016~\cite{toloudis2016associating} & \textit{Correlation analysis}: CVSS v2 \& WIVSS (0-10) & NVD & Principal component analysis \& Spearman correlation coefficient\\\hline
     
\end{tabular}
\end{table}

To provide even more fine-grained severity value than the categories, the last sub-theme has predicted the \textit{severity score} (see Table~\ref{tab:severity_score_studies}).
Using SV descriptions on NVD, Sahin et al.~\cite{sahin2019conceptual} compared the performance of ML-based regressors (e.g., XGBoost~\cite{chen2016xgboost} and Linear regression) and DL-based ones (e.g., CNN~\cite{kim2014convolutional} and LSTM~\cite{hochreiter1997long}) for predicting the severity score of CVSS version 2~\cite{cvss_v2}. These authors showed that DL-based approaches generally outperformed the ML-based counterparts. For CVSS version 3~\cite{cvss_v3,cvss_v31}, Chen et al.~\cite{chen2019vest,chen2019vase} and Anwar et al.~\cite{anwar2020cleaning} also reported the strong performance of DL-based models (e.g., CNN and graph convolutional neural network~\cite{kipf2016semi}). Some other studies did not directly predict severity score from SV descriptions, instead they aggregated the predicted values of the CVSS Exploitability (see section~\ref{sec:exploit_prediction}) and Impact metrics (see section~\ref{sec:impact_prediction}) using the formulas of CVSS version 2~\cite{wen2015novel,spanos2018multi,elbaz2020fighting,jiang2020approach}, version 3~\cite{ognawala2018automatically,elbaz2020fighting,jiang2020approach} and WIVSS~\cite{spanos2018multi}.
We noticed the papers predicting both versions (e.g., CVSS versions 2 vs. 3 or CVSS version 2 vs. WIVSS) usually obtained better performance for version 3 and WIVSS than version 2~\cite{elbaz2020fighting,jiang2020approach}. These findings may suggest that the improvements made by experts in version 3 and WIVSS compared to version 2 help make the patterns in severity score clearer and easier for ML models to capture.
In addition to predicting severity score, Toloudis et al.~\cite{toloudis2016associating} examined the correlation between words in descriptions of SVs and the severity values of such SVs, aiming to shed light on words that increase or decrease the severity score of SVs.

\subsection{Theme Discussion}
In the \textit{Severity} theme, predicting the severity levels is the most prevalent task, followed by severity score prediction and then binary classification of the severity.
In practice, severity score gives more fine-grained information (fewer SVs per value) for practitioners to rank/prioritize SVs than categorical/binary levels. However, predicting continuous score values is usually challenging and requires more robust models as this task involves higher uncertainty to learn inherent patterns from data than classifying fixed/discrete levels. We observed that DL models such as graph neural networks~\mbox{\cite{chen2019vest,chen2019vase}}, LSTM~\mbox{\cite{sahin2019conceptual}} and CNN~\mbox{\cite{anwar2020cleaning}} have been shown to be better than traditional ML models for predicting severity score. However, most of these studies did not evaluate their models in a continuous deployment setting to investigate how the models will cope with changing patterns of new SVs over time. We distill recommendations for future work on real-world application and evaluation of these models in section~\mbox{\ref{subsec:app_and_eval}}.

\section{Type Prediction}
\label{sec:type_prediction}

\begin{table}
\fontsize{8}{9}\selectfont
\centering
\caption{List of the surveyed papers in the \textit{Type} theme.}\label{tab:type_studies}
\vspace{-2pt}

\begin{tabular}{|p{1.5cm}|p{3.9cm}|p{2.3cm}|p{4.5cm}|}

 \hline \textbf{Study} & \textbf{Nature of task} & \textbf{Data source} & \textbf{Data-driven technique}\\\hline
 
    \multicolumn{4}{|c|}{\cellcolor[HTML]{C0C0C0}}\\
    \multicolumn{4}{|c|}{\multirow{-2}{*}{\cellcolor[HTML]{C0C0C0} \textbf{Sub-theme: 1. Common Weakness Enumeration (CWE)}}}\\\hline

    Wang et al. 2010~\cite{wang2010vulnerability} & \textit{Multi-class classification}: CWE classes & NVD, CVSS & Na\"ive Bayes\\\hhline{-~*{2}{-}}
    Shuai et al. 2013~\cite{shuai2013automatic} & & NVD & SVM\\\hhline{-~*{2}{-}}
    Na et al. 2016 ~\cite{na2016study} & & NVD & Na\"ive Bayes\\\hhline{-~*{2}{-}}
    Ruohonen et al. 2018~\cite{ruohonen2018toward} & & NVD, CWE, Snyk & tf-idf with 1/2/3-grams and cosine similarity\\\hhline{-~*{2}{-}}
    Huang et al. 2019~\cite{huang2019automatic} & & NVD, CWE & MLP, Linear SVM, Na\"ive Bayes, KNN\\\hhline{-~*{2}{-}}
    Aota et al. 2020~\cite{aota2020automation} & & NVD & Random forest, Linear SVM, Logistic regression, Decision tree, Extremely randomized trees, LGBM\\\hhline{-~*{2}{-}}
    Aghaei et al. 2020~\cite{aghaei2020threatzoom} & & NVD, CVE & Adaptive fully-connected neural network with one hidden layer\\\hhline{-~*{2}{-}}
    Das et al. 2021~\cite{das2021v2w} & & NVD, CWE & BERT, Deep Siamese network\\\hhline{-~*{2}{-}}
    \noalign{\vskip\doublerulesep\vskip-\arrayrulewidth}\hhline{-~*{2}{-}}
    
    Zou et al. 2019~\cite{zou2019muvuldeepecker} & & NVD \& Software Assurance Reference Dataset (SARD) & Three Bi-LSTM models for extracting and combining global and local features from code functions\\\hline\hline
      
    Murtaza et al. 2016~\cite{murtaza2016mining} & \textit{Unsupervised learning}: sequence mining of SV types (over time) & NVD (CWE \& CPE) & 2/3/4/5-grams of CWEs\\\hline
      
    Lin et al. 2017~\cite{lin2017machine} & \textit{Unsupervised learning}: association rule mining of CWE-related aspects (prog. language, time of introduction \& consequence scope) & CWE & FP-growth association rule mining algorithm\\\hline
      
     Han et al. 2018~\cite{han2018deepweak} & \textit{Binary/Multi-class classification}: CWE relationships (CWE links, link types \& CWE consequences) & CWE & Deep knowledge graph embedding of CWE entities\\\hline
    
    \multicolumn{4}{|c|}{\cellcolor[HTML]{C0C0C0}}\\
    \multicolumn{4}{|c|}{\multirow{-2}{*}{\cellcolor[HTML]{C0C0C0} \textbf{Sub-theme: 2. Custom Vulnerability Types}}}\\\hline
      
    Venter et al. 2008~\cite{venter2008standardising}  & \textit{Unsupervised learning}: clustering & CVE & Self-organizing map\\\hline
  
    Neuhaus et al. 2010~\cite{neuhaus2010security}  & \textit{Unsupervised learning}: topic modeling & CVE & Latent Dirichlet Allocation (LDA)\\\hhline{-~*{2}{-}}
    Mounika et al.~\cite{mounika2019analyzing,vanamala2020topic} & & CVE, Open Web Application Security Project (OWASP) & LDA\\\hhline{-~*{2}{-}}
    Aljedaani et al. 2020~\cite{aljedaani2020lda} & & SV reports (Chromium project) & LDA\\\hline\hline
      
    Williams et al.~\cite{williams2018analyzing,williams2020vulnerability} & \multirowcell{2}[0ex][l]{\textit{Multi-class classification}: manually\\coded SV types} & NVD & Supervised Topical Evolution Model \& Diffusion-based storytelling technique\\\hhline{-~*{2}{-}}
    Russo et al. 2019~\cite{russo2019summarizing} & & NVD & Bayesian network, J48 tree, Logistic regression, Na\"ive Bayes, Random forest\\\hhline{-~*{2}{-}}
    Yan et al. 2017~\cite{yan2017exploitmeter} & & Executables of 100 Linux applications & Decision tree\\\hline
    
    Zhang et al. 2020~\cite{zhang2020general} & \textit{Multi-class classification}: platform-specific vulnerability types & China National Vulnerability Database & Logistic regression, Linear discriminant analysis, KNN, CART, SVM, bagging/boosting models\\\hline
\end{tabular}
\end{table}
\vspace{-3pt}

This section reports the work done in the \textit{Type} theme. Type groups SVs with similar characteristics, e.g., causes, attack patterns and impacts, and thus facilitating the reuse of known prioritization and remediation strategies employed for prior SVs of the same types.
Two key prediction outputs are: (\textit{i}) Common Weakness Enumeration (CWE) and (\textit{ii}) Custom vulnerability types (see Table ~\ref{tab:type_studies}).

\subsection{Summary of Primary Studies}
\subsubsection{Common Weakness Enumeration (CWE)}\label{subsubsec:cwe}

The first sub-theme determines and analyzes the patterns of the SV types provided by \textit{CWE}~\cite{cwe}.
CWE is currently the standard for SV types with more than 900 entries.
The first group of studies has focused on multi-class classification of these CWEs. Wang et al.~\cite{wang2010vulnerability} were the first to tackle this problem with a Na\"ive Bayes model using the CVSS metrics (version 2)~\cite{cvss_v2} and product names. Later, Shuai et al.~\cite{shuai2013automatic} used LDA~\cite{blei2003latent} with a location-aware weighting to extract important features from SV descriptions for building an effective SVM-based CWE classifier. Na et al.~\cite{na2016study} also showed that features extracted from SV descriptions can improve the Na\"ive Bayes model in~\cite{wang2010vulnerability}.
Ruohonen et al.~\cite{ruohonen2018toward} studied an information retrieval method, i.e., term-frequency inverse document frequency (tf-idf) and cosine similarity, to detect the CWE-ID with a description most similar to that of a given SV collected from NVD and Snyk.\footnote{\url{https://snyk.io/vuln}} This method performed well for CWEs without clear patterns/keywords in SV descriptions.
Aota et al.~\cite{aota2020automation} utilized the Boruta feature selection algorithm~\cite{kursa2010boruta} and Random forest to improve the performance of \textit{base} CWE classification.
Base CWEs give more fine-grained information for SV remediation than categorical CWEs used in~\cite{na2016study}.

There has been a recent rise in using neural network/DL based models for CWE classification. Huang et al.~\cite{huang2019automatic} implemented a deep neural network with tf-idf and information gain for the task and obtained better performance than SVM, Na\"ive Bayes and KNN. Aghaei et al.~\cite{aghaei2020threatzoom} improved upon~\cite{aota2020automation} for both categorical (coarse-grained) and base (fine-grained) CWE classification with an adaptive hierarchical neural network to determine sequences of less to more fine-grained CWEs. To capture the hierarchical structure and rare classes of CWEs, Das et al.~\cite{das2021v2w} matched SV and CWE descriptions instead of predicting CWEs directly.
They presented a deep Siamese network with a BERT-based~\cite{devlin2018bert} shared feature extractor that outperformed many baselines even for rare/unseen CWE classes. Recently, Zou et al.~\cite{zou2019muvuldeepecker} pioneered the multi-class classification of CWE in vulnerable functions curated from Software Assurance Reference Dataset (SARD)~\cite{sard} and NVD. They achieved high performance ($\sim$95\% F1-score) with DL (Bi-LSTM) models. The strength of their model came from combining global (semantically related statements) and local (variables/statements affecting function calls) features. Note that this model currently only works for functions in C/C++ and 40 selected classes of CWE.

Another group of studies has considered unsupervised learning methods to extract CWE sequences, patterns and relationships.
Sequences of SV types over time were identified by Murtaza et al.~\cite{murtaza2016mining} using an $n$-gram model.
This model sheds light on both co-occurring and upcoming CWEs (grams), raising awareness of potential cascading attacks. Lin et al.~\cite{lin2017machine} applied an association rule mining algorithm, FP-growth~\cite{han2000mining}, to extract the rules/patterns of various CWEs aspects including types, programming language, time of introduction and consequence scope. For example, buffer overflow (CWE type) usually appears during the implementation phase (time of introduction) in C/C++ (programming language) and affects the availability (consequence scope). Lately, Han et al.~\cite{han2018deepweak} developed a deep knowledge graph embedding technique to mine the relationships among CWE types, assisting in finding relevant SV types with similar properties.

\subsubsection{Custom Vulnerability Types}\label{subsubsec:custom_types}

The second sub-theme is about \textit{custom vulnerability types} other than CWE.
Venter et al.~\cite{venter2008standardising} used Self-organizing map~\cite{kohonen1990self}, an unsupervised clustering algorithm, to group SVs with similar descriptions on CVE. This was one of the earliest studies that automated SV type classification.
Topic modeling is another popular unsupervised learning model~\cite{neuhaus2010security,mounika2019analyzing,vanamala2020topic,aljedaani2020lda} to categorize SVs without an existing taxonomy. Neuhaus et al.~\cite{neuhaus2010security} applied LDA~\cite{blei2003latent} on SV descriptions to identify 28 prevalent SV types and then analyzed the trends of such types over time. The identified SV topics/types had considerable overlaps (up to 98\% precision and 95\% recall) with CWEs. Mounika et al.~\cite{mounika2019analyzing,vanamala2020topic} extended~\cite{neuhaus2010security} to map the LDA topics with the top-10 OWASP~\cite{owasp_website}. However, the LDA topics/keywords did not agree well ($<$ 40\%) with the OWASP descriptions, probably because 10 topics did not cover all the underlying patterns of SV descriptions.
Aljedaani et al.~\cite{aljedaani2020lda} again used LDA to identify 10 types of SVs reported in the bug tracking system of Chromium\footnote{\url{https://bugs.chromium.org/p/chromium/issues/list}} and found memory-related issues were the most prevalent topics.

Another group of studies has classified manually defined/selected SV types rather than CWE as some SV types are encountered more often in practice and require more attention. Williams et al.~\cite{williams2018analyzing,williams2020vulnerability} applied a supervised topical evolution model~\cite{naim2017scalable} to identify the features that best described the 10 pre-defined SV types\footnote{1. Buffer errors, 2. Cross-site scripting, 3. Path traversal, 4. Permissions and Privileges, 5. Input validation, 6. SQL injection, 7. Information disclosure, 8. Resources Error, 9. Cryptographic issues, 10. Code injection.} prevalent in the wild.
These authors then used a diffusion-based storytelling technique~\cite{barranco2019analyzing} to show the evolution of a particular topic of SVs over time; e.g., increasing API-related SVs requires hardening the APIs used in a product. To support user-friendly SV assessment using ever-increasing unstructured SV data, Russo et al.~\cite{russo2019summarizing} used Bayesian network to predict 10 pre-defined SV types.\footnote{1. Authentication bypass or Improper Authorization, 2. Cross-site scripting or HTML injection, 3. Denial of service, 4. Directory Traversal, 5. Local/Remote file include and Arbitrary file upload, 6. Information disclosure and/or Arbitrary file read, 7. Buffer/stack/heap/integer overflow, 8. Remote code execution, 9. SQL injection, 10. Unspecified vulnerability}
Besides predicting manually defined SV types using SV natural language descriptions, Yan et al.~\cite{yan2017exploitmeter} used a decision tree to predict 22 SV types prevalent in the executables of Linux applications. The predicted type was then combined with fuzzers' outputs to predict SV exploitability (see section~\ref{subsubsec:exploit_likelihood}).
Besides author-defined types, custom SV types also come from specific SV platforms.
Zhang et al.~\cite{zhang2020general} designed an ML-based framework to predict the SV types collected from China National Vulnerability Database. Ensemble models (bagging and boosting models) achieved, on average, the highest performance for this task.

\subsection{Theme Discussion}
In the \textit{Type} theme, detecting and characterizing coarse-grained and fine-grained CWE-based SV types are the frequent tasks. The large number and hierarchical structure of classes are the main challenges with CWE classification/analysis. In terms of solutions, deep Siamese networks~\mbox{\cite{das2021v2w}} are more robust to the class imbalance issue (due to many CWE classes), while graph-based neural networks~\mbox{\cite{han2018deepweak}} can effectively capture the hierarchical structure of CWEs. Future work can investigate the combination of these two types of DL architectures to solve both issues simultaneously. We also recommend more solutions for addressing the data imbalance issue in section~\mbox{\ref{subsubsec:relax_supervised}}. Besides model-level solutions, author-selected or platform-specific SV types have been considered to reduce the complexity of CWE. However, similar to custom SV consequences in section~\mbox{\ref{subsubsec:other_consequences}}, there is not yet a universally accepted taxonomy for these custom SV types. To reduce the subjectivity in selecting SV types for prediction, we suggest that future work should focus on the types that are commonly encountered and discussed by developers in the wild (see section~\mbox{\ref{subsubsec:developer_sources}} for more details).

\vspace{-3pt}

\section{Miscellaneous Tasks}
\label{sec:miscellaneous}

The last theme is \textit{Miscellaneous Tasks} covering the studies that are representative yet do not fit into the four previous themes.
This theme has three main sub-themes/tasks: (\textit{i}) Vulnerability information retrieval, (\textit{ii}) Cross-source vulnerability patterns and (\textit{iii}) Vulnerability fixing effort (see Table ~\ref{tab:misc_studies}).

\vspace{-3pt}

\begin{table}
\fontsize{8}{9}\selectfont
\centering

\caption{List of the surveyed papers in the \textit{Miscellaneous Tasks} theme.}
\label{tab:misc_studies}

\begin{tabular}{|p{1.5cm}|p{4.7cm}|p{3cm}|p{3cm}|}

    \hline \textbf{Study} & \textbf{Nature of task} & \textbf{Data source} & \textbf{Data-driven technique}\\\hline
    
    \multicolumn{4}{|c|}{\cellcolor[HTML]{C0C0C0}}\\
    \multicolumn{4}{|c|}{\multirow{-2}{*}{\cellcolor[HTML]{C0C0C0} \textbf{Sub-theme: 1. Vulnerability Information Retrieval}}}\\\hline
    
    Weeraward-hana et al. 2014~\cite{weerawardhana2014automated} & \textit{Multi-class classification}: Extraction of entities (software name/version, impact, attacker/user actions) from SV descriptions & NVD (210 randomly selected and manually labeled SVs) & Stanford Named Entity Recognizer implementing a CRF classifier\\\hline
 
    Dong et al. 2019~\cite{dong2019towards} & \textit{Multi-class classification}: Vulnerable software names/versions & CVE Details, NVD, ExploitDB, SecurityFocus, SecurityFocus Forum, SecurityTracker, Openwall & Word-level and character-level Bi-LSTM with attention mechanism\\\hline
    
    Gonzalez et al. 2019~\cite{gonzalez2019automated} & \textit{Multi-class classification}: Extraction of 19 Vulnerability Description Ontology~\cite{vdo} classes from SV descriptions & NVD & Na\"ive Bayes, Decision tree, SVM, Random forest, Majority voting model\\\hline

    Binyamini et al. 2020~\cite{binyamini2021framework} & Multi-class classification: Extraction of entities (attack vector/means/technique, privilege, impact, vulnerable platform/version/OS, network protocol/port) from SV descriptions to generate MulVal~\cite{ou2005mulval} interaction rules & NVD & Bi-LSTM with various feature extractors: word2vec, ELMo, BERT (pre-trained or trained from scratch)\\\hline

    Guo et al. 2020~\cite{guo2020predicting} & \textit{Multi-class classification}: Extraction of entities (SV type, root cause, attack type, attack vector) from SV descriptions & NVD, SecurityFocus & CNN, Bi-LSTM (with or without attention mechanism)\\\hline
 
    Waareus et al. 2020~\cite{waareus2020automated} & \textit{Multi-class classification}: Common Product Enumeration (CPE) & NVD & Word-level and character-level Bi-LSTM\\\hline
 
    Yitagesu et al. 2021~\cite{yitagesu2021vulnpos} & \textit{Multi-class classification}: Part-of-speech tagging of SV descriptions & NVD, CVE, CWE, CAPEC, CPE, Twitter, PTB corpus~\cite{marcus1993building} & Bi-LSTM\\\hline
 
    Sun et al. 2021~\cite{sun2021generating} & \textit{Multi-class classification}: Extraction of entities (vulnerable product/version/component, type, attack type, root cause, attack vector, impact) from ExploitDB to generate SV descriptions & NVD, ExploitDB & BERT models\\\hline
    
    \multicolumn{4}{|c|}{\cellcolor[HTML]{C0C0C0}}\\
    \multicolumn{4}{|c|}{\multirow{-2}{*}{\cellcolor[HTML]{C0C0C0} \textbf{Sub-theme: 2. Cross-source Vulnerability Patterns}}}\\\hline

    Horawalavith-ana et al. 2019~\cite{horawalavithana2019mentions} & \textit{Regression}: Number of software development activities on GitHub after disclosure of SVs & Twitter, Reddit, GitHub & MLP, LSTM\\\hline
    
    Xiao et al. 2019~\cite{xiao2019embedding} & \textit{Knowledge-graph reasoning}: modeling the relationships among SVs, its types and attack patterns & CVE, CWE, CAPEC (Linux project) & Translation-based knowledge-graph embedding\\\hline
    
    \multicolumn{4}{|c|}{\cellcolor[HTML]{C0C0C0}}\\
    \multicolumn{4}{|c|}{\multirow{-2}{*}{\cellcolor[HTML]{C0C0C0} \textbf{Sub-theme: 3. Vulnerability Fixing Effort}}}\\\hline

    Othmane et al. 2017~\cite{othmane2017time} & \textit{Regression}: time (days) to fix SVs & Proprietary SV data collected at the SAP company & Linear/Tree-based/Neural network regression\\\hline

\end{tabular}
\end{table}

\subsection{Summary of Primary Studies}
\subsubsection{Vulnerability Information Retrieval}\label{subsec:information_retrieval}

The first and major sub-theme is \textit{vulnerability information retrieval} that studies data-driven methods to extract different SV-related entities (e.g., affected products/versions) and their relationships from SV data. The current sub-theme extracts assessment information appearing explicitly in SV data (e.g., SV descriptions on NVD) rather than predicting implicit properties as done in prior sub-themes. For instance, CWE-119, i.e., ``Improper Restriction of Write Operations within the Bounds of a Memory Buffer'', can be retrieved directly from CVE-2020-28022,
but not from CVE-2021-2122.\footnote{\url{https://nvd.nist.gov/vuln/detail/CVE-2020-28022} \& \url{https://nvd.nist.gov/vuln/detail/CVE-2021-21220}} The latter case requires techniques from section~\ref{subsubsec:cwe}.

Most of the retrieval methods in this sub-theme have been formulated under the multi-class classification setting. One of the earliest works was conducted by Weerawardhana et al.~\cite{weerawardhana2014automated}. This study extracted software names/versions, impacts and attacker's/user's action from SV descriptions on NVD using Stanford Named Entity Recognition (NER) technique, a.k.a. CRF classifier~\cite{finkel2005incorporating}. Later, Dong et al.~\cite{dong2019towards} proposed to use a word/character-level Bi-LSTM to improve the performance of extracting vulnerable software names and versions from SV descriptions available on NVD and other SV databases/advisories (e.g., CVE Details~\cite{cve_details}, ExploitDB~\cite{exploitdb}, SecurityFocus~\cite{securityfocus}, SecurityTracker~\cite{securitytracker} and Openwall~\cite{openwall}). Based on the extracted entities, these authors also highlighted the inconsistencies in vulnerable software names and versions across different SV sources. Besides version products/names of SVs, Gonzalez et al.~\cite{gonzalez2019automated} used a majority vote of different ML models (e.g., SVM and Random forest) to extract the 19 entities of Vulnerability Description Ontology (VDO)~\cite{vdo} from SV descriptions to check the consistency of these descriptions based on the guidelines of VDO.
Since 2020, there has been a trend in using DL models (e.g., Bi-LSTM, CNNs or BERT~\cite{devlin2018bert}/ELMo~\cite{peters2018deep}) to extract different information from SV descriptions including required elements for generating MulVal~\cite{ou2005mulval} attack rules~\cite{binyamini2021framework} or SV types/root cause, attack type/vector~\cite{guo2020predicting}, Common Product Enumeration (CPE)~\cite{cpe} for standardizing names of vulnerable vendors/products/versions~\cite{waareus2020automated}, part-of-speech~\cite{yitagesu2021vulnpos} and relevant entities (e.g., vulnerable products, attack type, root cause) from ExploitDB to generate SV descriptions~\cite{sun2021generating}. BERT models~\cite{devlin2018bert}, pre-trained on general text (e.g., Wikipedia pages~\cite{wikipedia} or PTB corpus~\cite{marcus1993building}) and fine-tuned on SV text, have also been increasingly used to address the data scarcity/imbalance for the retrieval tasks.

\subsubsection{Cross-source Vulnerability Patterns}\label{subsubsec:cross_source_patterns}

The second sub-theme, \textit{cross-source vulnerability patterns}, finds commonality and/or discovers latent relationships among SV sources to enrich information for SV assessment and prioritization. Horawalavithana et al.~\cite{horawalavithana2019mentions} found a positive correlation between development activities (e.g., push/pull requests and issues) on GitHub and SV mentions on Reddit\footnote{\url{https://reddit.com}} and Twitter. These authors then used DL models (MLP~\cite{hastie2009elements} and LSTM~\cite{hochreiter1997long}) to predict the appearance and sequence of development activities when SVs were mentioned on the two social media platforms. Xiao et al.~\cite{xiao2019embedding} applied a translation-based graph embedding method to encode and predict the relationships among different SVs and the respective attack patterns and types. This work~\cite{xiao2019embedding} was based on DeepWeak of Han et al.~\cite{han2018deepweak}, but it still belongs to this sub-theme as they provided a multi-dimensional view of SVs using three different sources (NVD~\cite{nvd}, CWE~\cite{cwe} and CAPEC~\cite{capec}).
Xiao et al.~\cite{xiao2019embedding} envisioned that their knowledge graph can be extended to incorporate the source code introducing/fixing SVs.

\subsubsection{Vulnerability Fixing Effort}\label{subsubsec:fixing_effort}

The last sub-theme is \textit{vulnerability fixing effort} that focuses on estimating SV fixing effort through proxies such as the SV fixing time, usually in days.
Othmane and the co-authors were among the first to approach this problem.
These authors first conducted a large-scale qualitative study at the SAP company and identified 65 important code-based, process-based and developer-based factors contributing to the SV fixing effort~\cite{ben2015factors}. Later, the same group of authors~\cite{othmane2017time} leveraged the identified factors in their prior qualitative study to develop various regression models such as linear regression, tree-based regression and neural network regression models, to predict time-to-fix SVs using the data collected at SAP. These authors found that code components containing detected SVs are more important for the prediction than SV types.

\subsection{Theme Discussion}
In the \textit{Miscellaneous Tasks} theme, the key focus is on retrieving SV-related entities and characteristics from SV descriptions. The retrieval tasks are usually formulated as Named Entity Recognition from SV descriptions. However, we observed that NVD descriptions do not follow a consistent template~\mbox{\cite{anwar2020cleaning}}, posing significant challenges in labeling the entities for retrieval. The affected versions and vendor/product names of SVs also contain inconsistencies~\mbox{\cite{dong2019towards,anwar2020cleaning}}, making the retrieval tasks difficult. We recommend that data normalization and cleaning should be performed before labeling entities and building respective retrieval models to ensure the reliability of results.

Besides information retrieval, other tasks such as linking multi-sources, extracting cross-source patterns or estimating fixing effort are also useful to obtain richer SV information for assessment and prioritization, yet these tasks are still in early stages. Linking multiple sources and their patterns is the first step towards building an SV knowledge graph to answer different queries regarding a particular SV (e.g., what systems are affected, exploitation status, how to fix, or what SVs are similar). In the future, such a knowledge graph can be extended to capture artifacts of SVs in emerging software types like AI-based systems (see section~\mbox{\ref{subsec:sv_ap_data_driven_systems}}). Moreover, to advance SV fixing effort prediction, future work can consider adapting/customizing the existing practices/techniques used to predict fixing effort for general bugs~\mbox{\cite{zhang2013predicting,akbarinasaji2018predicting}}.

\section{Analysis of Data-driven Approaches for Software Vulnerability Assessment and Prioritization}\label{sec:elements_analysis}

\begin{table}
\fontsize{5.5}{6}\selectfont
\centering

\caption{The frequent data sources, features, models, evaluation techniques and evaluation metrics used for the five identified SV assessment and prioritization themes. \textbf{Notes}: The values are organized based on their overall frequency across the five themes. For the Prediction Model and Evaluation Metric elements, the values are first organized by categories (ML then DL for Prediction Model and classification then regression for Evaluation Metric) and then by frequencies. k-CV stands for k-fold cross-validation.
The full list of values and their appearance frequencies for the five elements in the five themes can be found at~\mbox{\cite{supp_materials}}.}
\label{tab:data_driven_elements}

\begin{tabular}{|p{3cm}|p{4.55cm}|p{5.2cm}|}

    \hline \textbf{Source/Technique/Metric} & \textbf{Strengths} & \textbf{Weaknesses}\\\hline
    
    \multicolumn{3}{|c|}{\cellcolor[HTML]{C0C0C0}}\\
    \multicolumn{3}{|c|}{\multirow{-2}{*}{\cellcolor[HTML]{C0C0C0} \textbf{Element: Data Source}}}\\\hline
    
    \makecell[l]{NVD/CVE/CVE Details (deprecated\\ OSVDB)} &
    \makecell[l]{\tabitem Report expert-verified information (with CVE-ID)\\ \tabitem Contain CWE and CVSS entries for each SV\\ \tabitem Link to external sources (official fixes or vendors' info)} & \makecell[l]{\tabitem Missing/incomplete links to vulnerable code/fixes\\ \tabitem Inconsistencies due to human errors\\ \tabitem Delayed SV reporting and assignment of CVSS metrics}\\\hline
    
    ExploitDB & 
    \makecell[l]{\tabitem Report PoC exploits of SVs (with links to CVE-ID)} & \makecell[l]{\tabitem May not lead to real exploits in the wild}\\\hline
    
    \makecell[l]{Other security advisories (e.g.,\\ SecurityFocus, Symantec or X-Force)} & 
    \makecell[l]{\tabitem Report real-world exploits of SVs\\ \tabitem Cover a wide range of SVs (including ones w/o CVE-ID)} & \makecell[l]{\tabitem Some exploits may not have links to CVE entries for\\ mapping with other assessment metrics}\\\hline
    
    \makecell[l]{Informal sources (e.g., Twitter\\ and darkweb)} & 
    \makecell[l]{\tabitem Early reporting of SVs (maybe even earlier than NVD)\\ \tabitem Contain non-technical SV information (e.g., financial\\ damage or socio-technical challenges in addressing SVs)} & \makecell[l]{\tabitem Contain non-verified and even misleading information\\ \tabitem May cause adversarial attacks to assessment models}\\\hline
    
    \multicolumn{3}{|c|}{\cellcolor[HTML]{C0C0C0}}\\
    \multicolumn{3}{|c|}{\multirow{-2}{*}{\cellcolor[HTML]{C0C0C0} \textbf{Element: Model Feature}}}\\\hline
    
    BoW/tf-idf/n-grams & 
    \makecell[l]{\tabitem Simple to implement\\ \tabitem Strong baseline for text-based inputs (e.g., SV descrip-\\tions in security databases/advisories)} & \makecell[l]{\tabitem May suffer from vocabulary explosion (e.g., many new descrip-\\tion words for new SVs)\\ \tabitem No consideration of word context/order (maybe needed for\\ code-based SV analysis)\\ \tabitem Cannot handle Out-of-Vocabulary (OoV) words (can be resolved\\ with subwords~\cite{le2019automated})}\\\hline
    
    Word2vec & 
    \makecell[l]{\tabitem Capture nearby context of each word\\ \tabitem Can reuse existing pre-trained model(s)} & \makecell[l]{\tabitem Cannot handle OoV words (can be resolved with fastText~\cite{bojanowski2017enriching})\\ \tabitem No consideration of word order}\\\hline
    
    \makecell[l]{DL model end-to-end trainable\\ features} & 
    \makecell[l]{\tabitem Produce SV task-specific features} & \makecell[l]{\tabitem May not produce high-quality representation for tasks\\ with limited data (e.g., real-world exploit prediction)}\\\hline
    
    \makecell[l]{Bidirectional Encoder Representations\\ from Transformers (BERT)} & 
    \makecell[l]{\tabitem Capture contextual representation of text (i.e., the\\ feature vector of a word is specific to each input)\\ \tabitem Capture word order in an input\\ \tabitem Can handle OoV words} & \makecell[l]{\tabitem May require GPU to speed up feature inference\\ \tabitem May be too computationally expensive and require too\\ much data to train a strong model from scratch\\ \tabitem May require fine-tuning to work well for a source task}\\\hline
    
    \makecell[l]{Source/expert-defined metadata\\ features} & 
    \makecell[l]{\tabitem Lightweight\\ \tabitem Human interpretable for a task of interest} & \makecell[l]{\tabitem Require SV expertise to define relevant features\\ \tabitem Hard to generalize to new tasks}\\\hline
    
    \multicolumn{3}{|c|}{\cellcolor[HTML]{C0C0C0}}\\
    \multicolumn{3}{|c|}{\multirow{-2}{*}{\cellcolor[HTML]{C0C0C0} \textbf{Element: Prediction Model}}}\\\hline
    
    \makecell[l]{Single ML models (e.g., Linear SVM,\\ Logistic regression, Naïve Bayes)} & 
    \makecell[l]{\tabitem Simple to implement\\ \tabitem Efficient to (re-)train on large data (e.g., using the entire\\ NVD database)} & \makecell[l]{\tabitem May be prone to overfitting\\ \tabitem Usually do not perform as well as ensemble/DL models}\\\hline
    
    \makecell[l]{Ensemble ML models (e.g., Ran-\\dom forest, XGBoost, LGBM)} & 
    \makecell[l]{\tabitem Strong baseline (usually stronger than single models)\\ \tabitem Less prone to overfitting} & \makecell[l]{\tabitem Take longer to train than single models}\\\hline
    
    \makecell[l]{Latent Dirichlet Allocation (LDA --\\ topic modeling)} & 
    \makecell[l]{\tabitem Require no labeled data for training\\ \tabitem Can provide features for supervised learning models} & \makecell[l]{\tabitem Require SV expertise to manually label generated topics\\ \tabitem May generate human non-interpretable topics}\\\hline
    
    \makecell[l]{Deep Multi-Layer Perceptron\\ (MLP)} & 
    \makecell[l]{\tabitem Work readily with tabular data (e.g., manually defined\\ features or BoW/tf-idf/n-grams)} & \makecell[l]{\tabitem Perform comparably yet are more costly compared to ensemble\\ ML models\\ \tabitem Less effective for unstructured data (e.g., SV descriptions)}\\\hline
    
    \makecell[l]{Deep Convolutional Neural\\ Networks (CNN)} & 
    \makecell[l]{\tabitem Capture local and hierarchical patterns of inputs\\ \tabitem Usually perform better than MLP for text-based data} & \makecell[l]{\tabitem Cannot effectively capture sequential order of inputs (maybe\\ needed for code-based SV analysis)}\\\hline
    
    \makecell[l]{Deep recurrent neural networks\\ (e.g., LSTM or Bi-LSTM)} & 
    \makecell[l]{\tabitem Capture short-/long-term dependencies from inputs\\ \tabitem Usually perform better than MLP for text-based data} & \makecell[l]{\tabitem May suffer from the information bottleneck issue (can be\\ resolved with attention mechanism~\cite{bahdanau2014neural})\\ \tabitem Usually take longer to train than CNNs}\\\hline
    
    \makecell[l]{Deep graph neural networks\\ (e.g., Graph convolutional network)} & 
    \makecell[l]{\tabitem Capture directed relationships among multiple SV\\ entities and sources} & \makecell[l]{\tabitem Require graph-based inputs to work\\ \tabitem More computationally expensive than other DL models}\\\hline
    
    \makecell[l]{Deep transfer learning with\\ fine-tuning (e.g., BERT with task-\\specific classification layer(s))} & 
    \makecell[l]{\tabitem Can improve the performance for tasks with small data\\ (e.g., real-world exploit prediction)} & \makecell[l]{\tabitem Require target task to have similar nature as source task}\\\hline
    
    \makecell[l]{Deep constrastive learning\\ (e.g., Siamese neural networks)} & 
    \makecell[l]{\tabitem Can improve performance for tasks with small data\\ \tabitem Robust to class imbalance (e.g., CWE classes)} & \makecell[l]{\tabitem Computationally expensive (two inputs instead of one)\\ \tabitem Do not directly produce class-wise probabilities}\\\hline
    
    \makecell[l]{Deep multi-task learning} & 
    \makecell[l]{\tabitem Can share features for predicting multiple tasks (e.g.,\\ CVSS metrics) simultaneously\\ \tabitem Reduce training/maintenance cost} & \makecell[l]{\tabitem Require predicted tasks to be related\\ \tabitem Hard to tune the prediction/performance of individual tasks}\\\hline
    
    \multicolumn{3}{|c|}{\cellcolor[HTML]{C0C0C0}}\\
    \multicolumn{3}{|c|}{\multirow{-2}{*}{\cellcolor[HTML]{C0C0C0} \textbf{Element: Evaluation Technique}}}\\\hline
    
    \makecell[l]{Single k-CV without test} & 
    \makecell[l]{\tabitem Easy to implement\\ \tabitem Reduce the randomness of results with multiple folds} & \makecell[l]{\tabitem Do not have a separate test set for validating optimized models\\ (can be resolved with separate test set(s))\\ \tabitem Maybe infeasible for expensive DL models\\ \tabitem Use future data/SVs for training, maybe leading to biased results}\\\hline
    
    \makecell[l]{Single/multiple random train/test\\ with/without val (using val to tune\\ hyperparameters)} & 
    \makecell[l]{\tabitem Easy to implement\\ \tabitem Reduce the randomness of results (the multiple version)} & \makecell[l]{\tabitem May produce unstable results (the single version)\\ \tabitem Maybe infeasible for expensive DL models (the multiple version)\\ \tabitem Use future data/SVs for training, maybe leading to biased results}\\\hline
    
    \makecell[l]{Single/multiple time-based train/test\\ with/without val (using val to tune\\ hyperparameters)} & 
    \makecell[l]{\tabitem Consider the temporal properties of SVs, simulating the\\ realistic evaluation of ever-increasing SVs in practice\\ \tabitem Reduce the randomness of results (the multiple version)} & \makecell[l]{\tabitem Similar drawbacks for the single \& multiple versions as the\\ random counterparts\\ \tabitem May result in uneven/small splits (e.g., many SVs in a year)}\\\hline
    
    \multicolumn{3}{|c|}{\cellcolor[HTML]{C0C0C0}}\\
    \multicolumn{3}{|c|}{\multirow{-2}{*}{\cellcolor[HTML]{C0C0C0} \textbf{Element: Evaluation Metric}}}\\\hline
    
    \makecell[l]{F1-score/Precision/Recall\\ (classification)} & 
    \makecell[l]{\tabitem Suitable for imbalanced data (common in SV assessment\\ and prioritization tasks)} & \makecell[l]{\tabitem Do not consider True Negatives in a confusion matrix (can be\\ resolved with Matthews Correlation Coefficient (MCC))}\\\hline
    
    \makecell[l]{Accuracy (classification)} & 
    \makecell[l]{\tabitem Consider all the cells in a confusion matrix} & \makecell[l]{\tabitem Unsuitable for imbalanced data (can be resolved with MCC)}\\\hline
    
    \makecell[l]{Area Under the Curve (AUC)\\ (classification)} & 
    \makecell[l]{\tabitem Independent of prediction thresholds} & \makecell[l]{\tabitem May not represent real-world settings (i.e., as models in practice\\ mostly use fixed classification thresholds)\\ \tabitem ROC-AUC may not be suitable for imbalanced data (can be\\ resolved with Precision-Recall-AUC)}\\\hline
    
    \makecell[l]{Mean absolute (percentage) error/\\ Root mean squared error (regression)} & 
    \makecell[l]{\tabitem Show absolute performance of models} & \makecell[l]{\tabitem Maybe hard to interpret a value on its own without domain\\ knowledge (i.e., whether an error of $x$ is sufficiently effective)}\\\hline
    
    \makecell[l]{Correlation coefficient ($r$)/ Coef.\\ of determination ($R^2$) (regression)} & 
    \makecell[l]{\tabitem Show relative performance of models (0 -- 1), where 0 is\\ worst \& 1 is best} & \makecell[l]{\tabitem $R^2$ always increases when adding any new feature (can be\\ resolved with adjusted $R^2$)}\\\hline

\end{tabular}
\end{table}

We extract and analyze the five key elements for data-driven SV assessment and prioritization: (\textit{i}) Data sources, (\textit{ii}) Model features, (\textit{iii}) Prediction models, (\textit{iv}) Evaluation techniques and (\textit{v}) Evaluation metrics.
These elements correspond to the four main steps in building data-driven models: data collection (data sources), feature engineering (model features), model training (prediction models) and model evaluation (evaluation techniques/metrics)~\cite{han2011data,sabir2021machine}.
We present the most common practices for each element in Table~\mbox{\ref{tab:data_driven_elements}}.

\subsection{Data sources}\label{subsec:data_source}

Identifying and collecting rich and reliable SV-related data are the first tasks to build data-driven models for automating SV assessment and prioritization tasks. As shown in Table~\ref{tab:data_driven_elements}, a wide variety of data sources have been considered to accomplish the five identified themes.

Across the five themes, NVD~\cite{nvd} and CVE~\cite{cve} have been the most prevalently used data sources.
The popularity of NVD/CVE is mainly because they publish expert-verified SV information that can be used to develop prediction models.
Firstly, many studies have considered SV descriptions on NVD/CVE as model inputs. Secondly, the SV characteristics on NVD have been heavily used as assessment outputs in all the themes, e.g., CVSS Exploitability metrics for \textit{Exploitation}, CVSS Impact/Scope metrics for \textit{Impact}, CVSS severity score/levels for \textit{Severity}, CWE for \textit{Type}, CWE/CPE for \textit{Miscellaneous tasks}. Thirdly, external sources on NVD/CVE have enabled many studies to obtain richer SV information (e.g., exploitation availability/time~\cite{chen2019vest} or vulnerable code/crashes~\cite{yan2017exploitmeter,tripathi2017exniffer}) and extract relationships among multiple SV sources to develop a knowledge graph of SVs (e.g.,~\cite{han2018deepweak,xiao2019embedding}).
However, NVD/CVE still suffer from information inconsistencies~\cite{dong2019towards,anwar2020cleaning} and missing relevant external sources (e.g., SV fixing code)~\cite{hommersom2021automated}.
Such issues motivate future work to validate/clean NVD data and utilize more sources for code-based SV assessment and prioritization (see section~\ref{subsubsec:developer_sources}).

To enrich the SV information on NVD/CVE, many other security advisories and SV databases have been commonly leveraged by the reviewed studies, notably ExploitDB~\cite{exploitdb}, Symantec~\cite{symantec,symantec_threat}, SecurityFocus~\cite{securityfocus}, CVE Details~\cite{cve_details} and OSVDB. Most of these sources disclose PoC (ExploitDB and OSVDB) and/or real-world (Symantec and Security Focus) exploits. However, real-world exploits are much rarer and different compared to PoC ones~\cite{sabottke2015vulnerability,jacobs2020improving}.
It is recommended that future work should explore more data sources (other than the ones in Table~\ref{tab:exploit_studies_likelihood}) and better methods to retrieve real-world exploits (see section~\ref{subsec:data_avail}).
Additionally, CVE Details and OSVDB are SV databases like NVD yet with a few key differences. CVE Details explicitly monitors Exploit-DB entries that may be missed on NVD and provides a more user-friendly interface to view/search SVs. OSVDB also reports SVs that do not appear on NVD (without CVE-ID), but this site was discontinued in 2016.

Besides official/expert-verified data sources, we have seen an increasing interest in mining SV information from informal sources that also contain non-expert generated content such as social media (e.g., Twitter) and darkweb. Especially, Twitter has been widely used for predicting exploits as this platform has been shown to contain many SV disclosures even before official databases like NVD~\cite{sabottke2015vulnerability,chen2019vase}. Recently, darkweb forums/sites/markets have also gained traction as SV mentions on these sites have a strong correlation with their exploits in the wild~\cite{almukaynizi2017proactive,almukaynizi2019patch}. However, SV-related data on these informal sources are much noisier because they neither follow any pre-defined structure nor have any verification and they are even prone to fake news~\cite{sabottke2015vulnerability}. Thus, the data integrity of these sources should be checked,
potentially by checking the reputation of posters, to avoid inputting unreliable data to prediction models and potentially producing misleading findings.

\subsection{Model features}\label{subsec:model_feature}
Collected raw data need to be represented by suitable features for training prediction models.
There are three key types of feature representation methods in this area: term frequency (e.g., BoW, tf-idf and n-grams), DL learned features (e.g., BERT and word2vec) and source/expert-defined metadata (e.g., CVSS metrics and CPE on NVD or tweet properties on Twitter), as summarized in Table~\ref{tab:data_driven_elements}.

Regarding the term-frequency based methods, BoW has been the most popular one. Its popularity is probably because it is one of the simplest ways to extract features from natural language descriptions of SVs and directly compatible with popular ML models (e.g., Linear SVM, Logistic regression and Random forest) in section~\ref{subsec:prediction_model}. Besides plain term count/frequency, other studies have also considered different weighting mechanisms such as inverse document frequency weighting (tf-idf) or tf-igm~\cite{chen2016turning} inverse gravity moment weighting (tf-igm). Tf-igm has been shown to work better than BoW and tf-idf at classifying severity~\cite{kudjo2019improving,chen2020automatic}.
Future work is still needed to evaluate the applicability and generalizability of tf-igm for other SV assessment and prioritization tasks.

Recently, Neural Network (NN) or DL based features such as word2vec~\cite{mikolov2013distributed} and BERT~\cite{devlin2018bert} have been increasingly used to improve the performance of predicting CVSS exploitation/impact/severity metrics~\cite{han2017learning,gong2019joint}, CWE types~\cite{das2021v2w} and SV information retrieval~\cite{guo2020predicting,waareus2020automated}. Compared to BoW and its variants, NN and DL can extract more efficient and context-aware features from vast SV data~\cite{le2020deep}. NN/DL techniques rely on distributed representation to encode SV-related words using fixed-length vectors much smaller than a vocabulary size. Moreover, these techniques capture the sequential order and context (nearby words) to enable better SV-related text comprehension (e.g., SV vs. general \textit{exploit}). Importantly, these NN/DL learned features can be first trained in a non-SV domain with abundant data (e.g., Wikipedia pages~\cite{wikipedia}) and then transferred/fine-tuned in the SV domain to address limited/imbalanced SV data~\cite{yin2020apply}. The main concern with these sophisticated NN/DL features is their limited interpretability, which is an exciting research area (see section~\mbox{\ref{subsubsec:interpretability}}).

The metadata about SVs can also complement the missing information in descriptions or code for SV assessment and prioritization. For example, prediction of exploits and their characteristics have been enhanced using CVSS metrics~\cite{almukaynizi2019patch}, CPE~\cite{aksu2018automated} and SV types~\cite{bhatt2021exploitability} on NVD. Additionally, Twitter-related statistics (e.g., number of followers, likes and retweets) have been shown to increase the performance of predicting SV exploitation, impact and severity~\cite{sabottke2015vulnerability,chen2019vest}. Recently, alongside features extracted from vulnerable code, the information about a software development process and involved developers have also been extracted to predict SV fixing effort~\cite{othmane2017time}. Currently, metadata-based and text-based features have been mainly integrated by concatenating their respective feature vectors (e.g.,~\mbox{\cite{chen2019vase,chen2019using,almukaynizi2017proactive,almukaynizi2019patch}}). An alternative yet unexplored way is to build separate models for each feature type and then combine these models using meta-learning (e.g., model stacking~\mbox{\cite{dzeroski2002combining}}).


\subsection{Prediction models}\label{subsec:prediction_model}

The extracted features enter a wide variety of ML/DL-based prediction models shown in Table~\ref{tab:data_driven_elements} to automate various SV assessment and prioritization tasks.
Classification techniques have the largest proportion, while regression and unsupervised techniques are less common.

Linear SVM~\cite{cortes1995support} has been the most frequently used classifier, especially in the Exploitation, Impact and Severity themes.
This popularity is expected as Linear SVM works well with the commonly used features, i.e., BoW and tf-idf.
Besides Linear SVM, Random forest, Na\"ive Bayes and Logistic regression have also been common classification models.
In recent years, advanced boosting models (e.g., XGBoost~\cite{chen2016xgboost} and LGBM~\cite{ke2017lightgbm}), and more lately, DL techniques (e.g., CNN~\cite{kim2014convolutional} and (Bi-)LSTM with attention~\cite{bahdanau2014neural}) have been increasingly utilized and shown better results than simple ML models like Linear SVM or Logistic regression.
In this area, some DL models are essential for certain tasks, e.g., building SV knowledge graph from multiple sources with graph neural networks~\mbox{\cite{kipf2016semi}}. DL models also offer solutions to data-related issues such as addressing class imbalance (e.g., deep Siamese network~\mbox{\cite{reimers2019sentence}}) or improving data efficiency (e.g., deep multi-task learning~\mbox{\cite{zhang2021survey}}).
Whenever applicable, it is recommended that future work should still consider simple baselines alongside sophisticated ones as simple methods can perform on par with advanced ones~\mbox{\cite{mazuera2021shallow}}.

Besides classification, various prediction models have also been investigated for regression (e.g., predicting exploit time, severity score and fixing time).
Linear SVM has again been the most commonly used regressor as SV descriptions have usually been the regression input.
Notably, many studies in the Severity theme did not build regression models to directly obtain the severity score (e.g.,~\cite{wen2015novel,ognawala2018automatically,elbaz2020fighting,jiang2020approach,spanos2018multi}). Instead, they used the formulas defined by assessment frameworks (e.g., CVSS versions 2/3~\cite{cvss_v2,cvss_v3} or WIVSS~\cite{spanos2013wivss}) to compute the severity score from the base metrics predicted by respective classification models.
We argue that more effort should be invested in determining the severity score directly from SV data as these severity formulas can be subjective~\mbox{\cite{spring2021time}}. We also observe that there is still limited use of DL models for regression compared to classification.

In addition to supervised (classification/regression) techniques, unsupervised learning has also been considered for extracting underlying patterns of SV data, especially in the Type theme. Latent Dirichlet Allocation (LDA)~\cite{blei2003latent} has been the most commonly used topic model to identify latent topics/types of SVs without relying on a labeled taxonomy. The identified topics were mapped to the existing SV taxonomies such as CWE~\cite{neuhaus2010security} and OWASP~\cite{mounika2019analyzing,vanamala2020topic}.
The topics generated by topic models like LDA can also be used as features for classification/regression models~\cite{ruohonen2017classifying} or building topic-wise models to capture local SV patterns~\cite{menzies2018500+}.
However, definite interpretations for unsupervised outputs are challenging to obtain as they usually rely on human judgement~\mbox{\cite{palacio2019evaluation}}.

\subsection{Evaluation techniques}\label{subsec:evaluation_technique}

It is important to evaluate a trained model to ensure the model meets certain requirements (e.g., advancing the state-of-the-art).
The evaluation generally needs to be conducted on a different set of data other than the training set to avoid overfitting and objectively estimate model generalizability~\cite{hastie2009elements}. The commonly used evaluation techniques are summarized in Table~\ref{tab:data_driven_elements}.

The reviewed studies have mostly used one or multiple validation and/or test sets\footnote{Validation set(s) helps optimize/tune a model (finding the best task/data-specific hyperparameters), and test set(s) evaluates the optimized/tuned model. Using only validation set(s) means evaluating a model with default/pre-defined hyperparameters.} to evaluate their models, in which each validation/test set has been either randomly or time-based selected. Specifically, k-fold cross-validation has been one of the most commonly used techniques. The number of folds has usually been 5 or 10, but less standard values like 4~\cite{yan2017exploitmeter} have also been used. However, k-fold cross-validation uses all parts of data at least once for training; thus, there is no hidden test set to evaluate the optimal model with the highest (cross-)validation performance.

To address the lack of hidden test set(s), a common practice in the studied papers has been to split a dataset into single training and test sets, sometimes with an additional validation set for tuning hyperparameters to obtain an optimal model. Recently, data has been increasingly split based on the published time of SVs to better reflect the changing nature of ever-increasing SVs~\cite{bullough2017predicting,le2019automated}. There have been various ratios for random (e.g., 80:20, 75:25 or 67:33) and time-based (e.g., week/month/year-wise) splits. However, the results reported using single validation/test sets may be unstable (i.e., unreproducible results using different set(s))~\cite{raschka2018model}.

To ensure both the time order and reduce the result randomness, we recommend using multiple splits of training and test sets in combination with time-based validation in each training set. Statistical analyses (e.g., hypothesis testing and effect size) should also be conducted to confirm the reliability of findings with respect to the randomization of models/data in multiple runs~\cite{de2019evolution}.

\subsection{Evaluation metrics}\label{subsec:evaluation_metric}

Evaluating different aspects of a model requires respective proper metrics.
The popular metrics for evaluating the tasks in each theme are given in Table~\ref{tab:data_driven_elements}.

Across the five themes, Accuracy, Precision, Recall and F1-score~\cite{jiao2016performance} have been the most commonly used metrics because of a large number of classification tasks in the five themes.
However, Accuracy is not a suitable measure for SV assessment and prioritization tasks with imbalanced data (e.g., SVs with real-world exploits vs. non-exploited SVs).
The sample size of one class is much smaller than the others, and thus the overall Accuracy would be dominated by the majority classes.
Besides these four commonly used metrics, AUC based on the ROC curve (ROC-AUC)~\cite{jiao2016performance} has also been considered as it is threshold-independent. However, we suggest that ROC-AUC should be used with caution in practice as most deployed models would have a fixed decision threshold (e.g., 0.5). Instead of ROC-AUC, we suggest Matthews Correlation Coefficient~\cite{jiao2016performance} (MCC) as a more meaningful evaluation metric to be considered as it explicitly captures all values in a confusion matrix, and thus has less bias in results.

For regression tasks, various metrics have been used such as Mean absolute error, Mean absolute percentage error, Root mean squared error~\cite{spanos2018multi} as well as Correlation coefficient ($r$) and Coefficient of determination ($R^2$)~\cite{othmane2017time}. Note that \textit{adjusted} $R^2$ should be preferred over $R^2$ as $R^2$ would always increase when adding a new (even irrelevant) feature.

A model can have a higher value of one metric yet lower values of others.\footnote{\url{https://stackoverflow.com/questions/34698161}} Therefore, we suggest using a combination of suitable metrics for a task of interest to avoid result bias towards a specific metric. Currently, most studies have focused on evaluating model effectiveness, i.e., how well the predicted outputs match the ground-truth values. Besides effectiveness, other aspects (e.g., efficiency in training/deployment and robustness to input changes) of models should also be evaluated to provide a complete picture of model applicability in practice (see section~\ref{subsubsec:realistic_eval}).

\section{Open Research Challenges and Future Directions of Data-driven Software Vulnerability Assessment and Prioritization}

\begin{figure*}[t]

  \centering
  \includegraphics[width=0.9\linewidth,keepaspectratio]{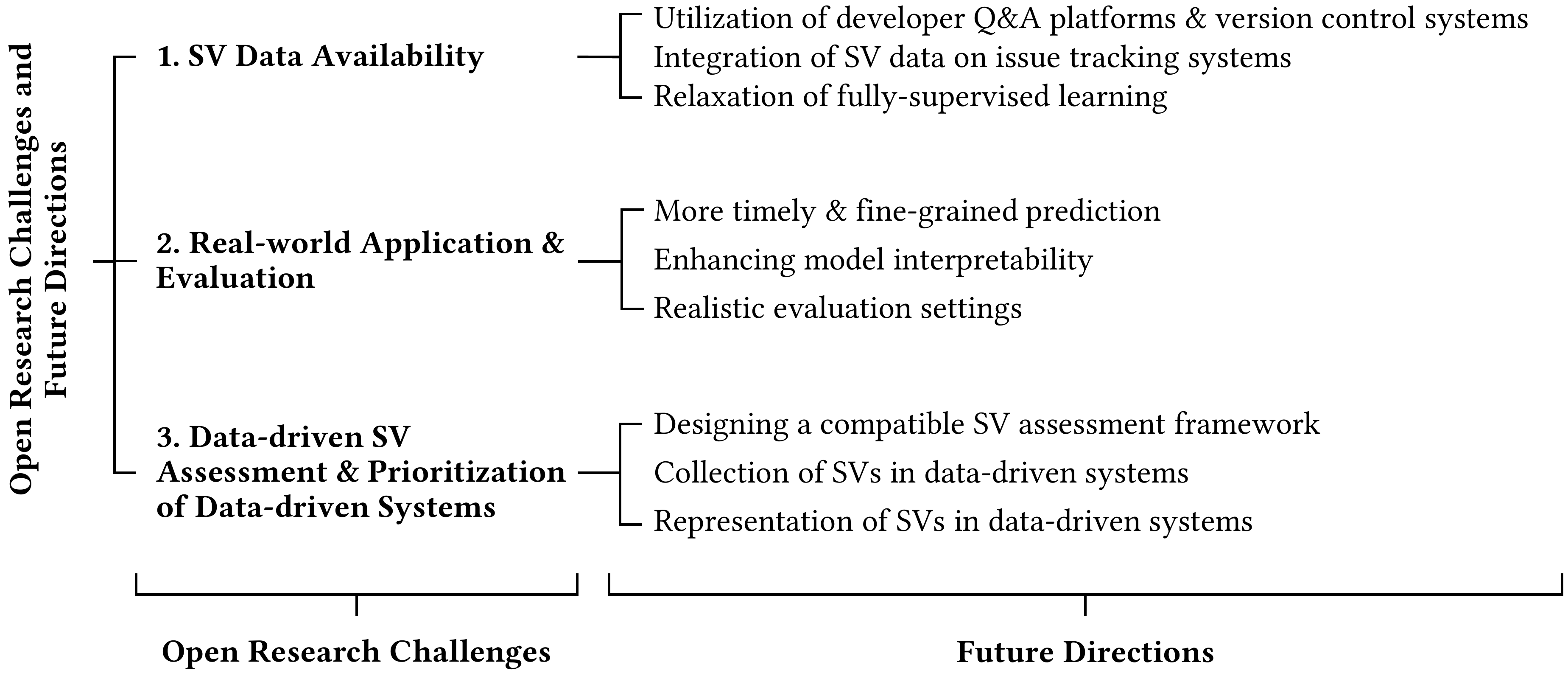}

  \caption{List of challenges and future directions for data-driven SV assessment and prioritization.}
  \label{fig:challenges_future}
\end{figure*}

We discuss three main open challenges with the reviewed studies of data-driven SV assessment and prioritization and then present nine potential directions to address such challenges (see Figure~\ref{fig:challenges_future}).

\label{sec:challenges}

\subsection{SV Data Availability}\label{subsec:data_avail}

This section focuses on three key issues with the currently used data sources and potential solutions.
First, the current data hardly contain specific developer's concerns and practices when addressing real-world SVs (section~\ref{subsubsec:developer_sources}). Second, the data sources still miss many SV-related bugs reported in issue tracking systems, limiting the amount of data for training prediction models (section~\ref{subsubsec:bug_vs_sv}). Third, some tasks/outputs (e.g., real-world exploit prediction) suffer from limited and/or imbalanced labeled data, potentially leading to unreliable performance of fully-supervised models (section~\ref{subsubsec:relax_supervised}).

\subsubsection{Utilization of developer Q\&A platforms and version control systems}\label{subsubsec:developer_sources}

\textit{Developer Question \& Answer (Q\&A) platforms} like Stack Overflow and Security StackExchange\footnote{Stack Overflow: \url{https://stackoverflow.com} \& Security StackExchange: \url{https://security.stackexchange.com}} contain tens of thousands of posts about challenges and solutions shared by millions of developers when tackling known SVs in real-world scenarios~\cite{le2021large}.
One of the key insights of Le et al.~\cite{le2021large}'s study is that the top SV types that developers usually struggle with are not always the same as those reported on SV databases (CWE~\cite{cwe} or OWASP~\cite{owasp_website}). Thus, future work should also consider real-world development-related issues discussed on developer Q\&A platforms for automatically assessing and prioritizing SVs. For example, the fixing effort of SVs may depend on the technical difficulty of implementing the respective mitigation strategies in a language or system of interest.

\textit{Version control systems} like GitHub\footnote{\url{https://github.com}} provide details about how developers addressed past SVs in real-world projects. Shrestha et al.~\cite{shrestha2020multiple} found developers sometimes discuss/disclose SV-related information on GitHub discussions even before the studied social media such as Twitter or Reddit. These findings show the potential of using GitHub discussions to complement the current sources for earlier SV assessment and prioritization. GitHub can also provide vulnerable code for performing assessment and prioritization for SVs rooted in source code~\cite{le2022use}, which is important yet has received limited attention from the community so far.
Moreover, Walden~\cite{walden2020impact} demonstrated the impact of a major SV (i.e., Heartbleed) on the characteristics (e.g., code complexity/style, contributors and development practices) of a single project (i.e., OpenSSL). Based on Walden's findings, future work can study whether the impact of an SV would be similar or different in multiple affected projects. Such investigation would give insights into the possibility of leveraging data from large projects to perform SV assessment and prioritization in smaller projects with the same/similar SVs.

\subsubsection{Integration of SV data on issue tracking systems}\label{subsubsec:bug_vs_sv}

\textit{Issue/bug tracking systems} like JIRA, Bugzilla or GitHub issues\footnote{JIRA: \url{https://www.atlassian.com/software/jira}, Bugzilla: \url{https://www.bugzilla.org/}, GitHub issues: \url{https://docs.github.com/en/issues}} have been reporting numerous security-related bugs, many of which are SVs, but they have been underexplored for data-driven SV assessment and prioritization.
Besides providing SV descriptions like CVE/NVD, these bug reports also contain other artifacts such as steps to reproduce, stack traces and test cases that give extra information about SVs~\cite{zimmermann2010makes}.
However, it is not trivial to obtain and integrate these SV-related bug reports with the ones on SV databases.

One way to retrieve SVs on issue tracking systems is to use security bug reports~\cite{bhuiyan2021security}. Much research work has been put into developing effective models to automatically retrieve security bug reports (e.g.,~\cite{gegick2010identifying,peters2017text,wu2021data}). Among these studies, Wu et al.~\cite{wu2021data} manually verified and cleaned the security bug reports to provide a clean dataset for automated security bug report identification. However, more of such manual effort is still required to obtain up-to-date data because the original security bug reports in~\cite{wu2021data} were actually a part of the dataset collected back in 2014~\cite{ohira2015dataset}.

It is worth noting that \textit{not} all security bug reports are related to SVs such as issues/improvements in implementing security features.\footnote{The security bug report AMBARI-1373 on JIRA (\url{https://issues.apache.org/jira/browse/AMBARI-1373}) was about improving the front-end of AMBARI Web by displaying the current logged in user.} Thus, future studies need to filter out these cases before using security bug reports for SV assessment and prioritization.
We also emphasize that some SV-related bug reports are overlapping with the ones on NVD (e.g., the SV report AMBARI-14780\footnote{\url{https://issues.apache.org/jira/browse/AMBARI-14780}} on JIRA refers to CVE-2016-0731 on CVE/NVD). Such overlaps would require data cleaning during the integration of reports on issue tracking systems and SV databases to avoid data duplication (e.g., similar SV descriptions) when developing SV assessment and prioritization models.

\subsubsection{Relaxation of fully-supervised learning}\label{subsubsec:relax_supervised}

Supervised learning models of many tasks in the five themes (see section~\ref{subsec:prediction_model}) require fully labeled data, but the data of some tasks are quite limited. To address the data-hungriness of these fully-supervised learning models, future studies can approach the SV assessment and prioritization tasks with \textit{low-shot learning} and/or \textit{semi-supervised learning}.

\textit{Low-shot learning} a.k.a. \textit{few-shot learning} is designed to perform supervised learning using only a few examples per class, significantly reducing the labeling effort~\cite{wang2020generalizing}.
So far, only one study utilized low-shot learning with a deep Siamese network~\cite{das2021v2w} (i.e., a shared feature model with similarity learning) to effectively predict SV types (CWE) and even generalize to unseen classes (i.e., zero-shot learning). There are still many opportunities for investigating different few-shot learning techniques for other SV assessment and prioritization tasks. Note that the shared features in few-shot learning can also be enhanced with pretrained models (e.g., BERT~\cite{devlin2018bert}) on another domain/task/project with more labeled data than the current task/project in the SV domain.

\textit{Semi-supervised learning} enables training models with limited labeled data yet a large amount of unlabeled data~\cite{van2020survey}, potentially leveraging hidden/unlabeled SVs in the wild. Recently, we have seen an increasing interest in using different techniques of this learning paradigm in the SV domain such as collecting SV patches using multi-view co-training~\cite{sawadogo2020learning} or retrieving SV discussions on developer Q\&A sites using positive-unlabeled learning~\cite{le2020puminer}. However, it is still little known about the effectiveness of semi-supervised learning for SV assessment and prioritization.

\subsection{Real-world Application and Evaluation}\label{subsec:app_and_eval}

The \textit{experimental} performance of some SV assessment and prioritization models is promising, but the \textit{real-world} applicability of such models is still questionable. First, these models may not be useful in practice due to delayed inputs and coarse-grained outputs (section~\ref{subsubsec:timely_finegrained_prediction}). Second, many models are black-box, limiting the understanding of the model predictions (section~\ref{subsubsec:interpretability}). Third, some models are evaluated in over-optimistic conditions far from real-world scenarios (section~\ref{subsubsec:realistic_eval}).

\subsubsection{More timely and fine-grained prediction}\label{subsubsec:timely_finegrained_prediction}

Although SV descriptions have been commonly used as model inputs (see section~\ref{subsec:data_source}), these descriptions are usually published long after SVs introduced/discovered in codebases~\cite{meneely2013patch}. One potential solution to this issue is to perform assessment and prioritization of SVs in code commits. Code commits contain changes made by developers to fix a bug/SV, implement a new feature or refactor code, and new SVs may be introduced in such changes~\cite{bosu2014identifying}. Commit-level prediction would allow just-in-time SV assessment and prioritization as soon as SVs are introduced, reducing the waiting time for SV information to be verified and published on security advisories/databases~\cite{le2021deepcva}. It should be noted that report-level prediction is still important for assessing and prioritizing third-party libraries/software, especially the ones without available code (commits), and/or SVs missed by commit-level prediction.

CVSS~\cite{cvss} has been most frequently used for assessing the exploitability, impact and severity levels/score of SVs (see sections~\ref{sec:exploit_prediction},~\ref{sec:impact_prediction} and~\ref{sec:severity_prediction}), but there are increasing concerns that CVSS outputs are still generic. Specifically, Spring et al.~\cite{spring2021time} argued that CVSS tends to provide one-size-fits-all assessment metrics regardless of the context of SVs; i.e., the same SVs in different domains/environments are assigned the same metric values. For instance, banking systems may consider the confidentiality and integrity of databases more important than the availability of web/app interfaces. In the future, alongside CVSS, prediction models should also incorporate the domain/business knowledge to customize the assessment of SVs to a system of interest (e.g., the impact of SVs on critical component(s) and/or the readiness of developers/solutions for mitigating such SVs in the current system). Future case studies with practitioners are also desired to correlate the quantitative performance of models and their usability/usefulness in real-world systems (e.g., reducing more critical SVs yet using fewer resources).

\subsubsection{Enhancing model interpretability}\label{subsubsec:interpretability}

Model interpretability is important to increase the transparency of the predictions made by a model, allowing practitioners to adjust the model/data to meet certain requirements~\cite{zhang2020survey}. Unfortunately, very few reviewed papers (e.g.,~\cite{toloudis2016associating,han2017learning}) explicitly discussed important features and/or explained why/when their models worked/failed for a task.

SV assessment and prioritization can draw inspiration from the related SV detection area where the interpretability of (DL-based) prediction models has been actively explored mainly by using (\textit{i}) specific model architectures/parameters or (\textit{ii}) external interpretation models/techniques~\cite{zhang2020survey}. In the first approach, prior studies successfully used the feature activation maps in a CNN model~\cite{russell2018automated} or leveraged attention-based neural network~\cite{duan2019vulsniper} to highlight and visualize the important code tokens that contribute to SVs. The second approach uses separate interpretation models on top of trained SV detectors. The interpretation models are either domain/model-agnostic~\cite{warnecke2020evaluating}, domain-agnostic yet specific to a model type (graph neural network~\cite{li2021vulnerability}) or SV-specific~\cite{zou2021interpreting}. The aforementioned approaches produce local/sample-wise interpretation, which can be aggregated to obtain global/task-wise interpretation. The global interpretation is similar to the feature importance of traditional ML models~\cite{chandrashekar2014survey} such as the weights of linear models (e.g., Logistic regression) or the (im)purity of nodes split by each feature in tree-based models (e.g., Random forest). However, it is still unclear about the applicability/effectiveness of these approaches for interpreting ML/DL-based SV assessment and prioritization models, requiring further investigations.

\vspace{-3pt}

\subsubsection{Realistic evaluation settings}\label{subsubsec:realistic_eval}

Most of the reviewed studies have evaluated their prediction models without capturing many factors encountered during the deployment of such models to production.
Specifically, the models used in practice would require to handle new data and be robust against adversarial data from informal sources such as social media or darkweb.

There are concerns with both predicting and integrating new SV data.
Regarding the prediction, Out-of-Vocabulary words in new data need to be properly accommodated to avoid performance degradation of prediction models~\cite{le2019automated}.
Regarding the new data integration, online/incremental training on new data can be considered instead of batch training on the whole dataset to reduce computational cost~\cite{cabral2019class}. The time-based splits should be used rather than random splits for evaluating online training to avoid leaking unseen (future) patterns to the model training (see section~\ref{subsec:evaluation_technique}).

Regarding the model robustness, only three reviewed studies considered adversarial attacks as part of their evaluation~\cite{sabottke2015vulnerability,xiao2018patching,almukaynizi2019patch}. However, a recent survey shows the prevalence of adversarial attacks targeted models in cybersecurity~\cite{rosenberg2021adversarial}. Thus, there is certainly a need for more evaluation of adversarial robustness for SV assessment and prioritization models, especially DL-based ones.

\vspace{-3pt}

\subsection{Data-driven SV Assessment and Prioritization of Data-driven Systems}\label{subsec:sv_ap_data_driven_systems}

Compared to other systems, reporting/analyzing SVs of data-driven/Artificial Intelligence (AI)-based systems is still in its infancy~\cite{kumar2020legal}.
Data-driven systems (e.g., smart recommender systems, chatbots, robots, and autonomous cars) are an emerging breed of systems whose cores are powered by AI technologies, e.g., ML and DL models built on data, rather than human-defined instructions as in traditional systems.
We discuss three key challenges of SV assessment and prioritization of data-driven systems compared to traditional systems and suggest potential solutions. Firstly, the current SV assessment frameworks need customizations to better reflect the nature of SVs in data-driven systems (section~\ref{subsubsec:new_assessment_framework}). Secondly, there is a lack of SVs collected from real-world data-driven systems, limiting the potential of data-driven SV assessment and prioritization (section~\ref{subsubsec:svs_datadriven_systems}). Thirdly, the current models require redesign, especially in the SV representation, to capture unique characteristics and artifacts of data-driven systems (section~\ref{subsubsec:datadriven_sv_representation}).

\vspace{-3pt}

\subsubsection{Designing a compatible SV assessment framework}\label{subsubsec:new_assessment_framework}
CVSS~\cite{cvss} is currently the most popular SV assessment framework for traditional systems, but its compatibility with data-driven systems still requires more investigation. The current CVSS documentation lacks instructions on how to assign metrics/score for SVs in data-driven systems. For example, it is unclear how to assign static CVSS metrics to systems with automatically updated data-driven models~\cite{chen2020machine} because adversarial examples for exploitation would likely change after the models are updated.
Such ambiguities should be clarified/resolved in future CVSS versions as data-driven systems become more prevalent.

The types of SVs in ML/DL models in data-driven systems are also mostly different from the ones provided by CWE~\cite{cwe}. The difference is mainly because these new SVs do not only emerge from configurations/code as in traditional systems, but also from training data and/or trained models~\cite{rosenberg2021adversarial}. Thus, we recommend that a new category of these SVs should be studied and potentially incorporated into CWE, similar to the newly added category for architectural SVs.\footnote{\url{https://cwe.mitre.org/data/definitions/1008.html}}

\subsubsection{Collection of SVs in data-driven systems}\label{subsubsec:svs_datadriven_systems}
To the best of our knowledge, there has been no existing large-scale dataset of SVs in ML/DL models deployed in real-world data-driven systems. Very few of such SVs have been reported in the wild, one of which is CVE-2019-20634.\footnote{\url{https://nvd.nist.gov/vuln/detail/CVE-2019-20634}} More of these SVs are required to help develop sufficiently effective SV assessment and prioritization models. One potential way to build such a dataset is to first match the (pre-trained) ML/DL models proposed in the literature or released on model repositories (e.g., Tensorflow Hub\footnote{\url{https://www.tensorflow.org/hub}}) with the ones used in real-world systems either on version control systems or in mobile apps~\cite{huang2021robustness}. The matched models would then be tested against known adversarial attacks to identify corresponding SVs. Notably, significant effort is still required to define/label assessment outputs of these SVs (see section~\ref{subsubsec:new_assessment_framework}).

\subsubsection{Representation of SVs in data-driven systems}\label{subsubsec:datadriven_sv_representation} Existing SV assessment and prioritization models for traditional systems have not considered unique data/model-related characteristics/features of data-driven systems~\cite{wan2019does}. Specifically, data-driven systems also encompass information about data (e.g., format, type, size and distribution) and ML/DL model(s) (e.g., configurations, parameters and performance). It is worth noting that SVs of ML/DL models in data-driven systems can also come from the frameworks used to develop such models (e.g., Tensorflow or Keras)\footnote{Tensorflow: \url{https://github.com/tensorflow/tensorflow} \& Keras: \url{https://github.com/keras-team/keras}}. However, developers of data-driven systems may not be aware of the (security) issues in the used ML/DL frameworks~\cite{liu2020using}. Thus, besides currently used features, future work should also consider the information about underlying data/models and used ML/DL development frameworks to improve the SV representation for building models to assess and prioritize SVs in data-driven systems.

\vspace{-3pt}

\section{Conclusions}
\label{sec:conclusions}

Assessment and prioritization are crucial phases to optimize resource utilization in addressing SVs at scale. The two phases have witnessed radical transformations following the increasing availability of SV data from multiple sources and advances in data-driven techniques. We presented a taxonomy to summarize the five main directions of the research work so far in this area. We identified and analyzed the key practices to develop data-driven models in the reviewed studies. We also highlighted the open challenges and suggested respective solutions to advance the field.

We envision the field will largely continue to improve the effectiveness of the presented tasks by leveraging more enriched data sources and sophisticated data-driven models, especially DL-based ones. Besides the improved performance, we also see many open opportunities/concerns in under-explored aspects of developing such advanced models. Overall, a deeper understanding of practitioners' concerns and real-world usage scenarios is the key to bridging the current gap between model development in academia and model deployment in production.

\vspace{-3pt}

\section*{Acknowledgment}
The work has been supported by the Cyber Security Research Centre Limited whose activities are partially funded by the Australian Government’s Cooperative Research Centres Programme.

\vspace{-3pt}

\bibliographystyle{ACM-Reference-Format}
\bibliography{reference}

\end{document}